\documentclass{emulateapj}

\newcommand{\tmin}{{T_{\rm min}}}
\newcommand{\msun}{M_{\odot}}
\renewcommand\ion[2]{#1$\;${\small\rmfamily{#2}}\relax}

\begin{document}

\title{The Tail of the Stripped Gas that Cooled: \ion{H}{I}, H$\alpha$ and X-ray Observational Signatures of Ram Pressure Stripping}
\author{Stephanie Tonnesen and Greg L. Bryan}
\affil{Department of Astronomy, Columbia University, Pupin Physics Laboratories, New York, NY 10027}

\begin{abstract}

Galaxies moving through the intracluster medium (ICM) of a cluster of galaxies can lose gas via ram pressure stripping.  This stripped gas forms a tail behind the galaxy which is potentially observable.  In this paper, we carry out hydrodynamical simulations of a galaxy undergoing stripping with a focus on the gas properties in the wake and their observational signatures.  We include radiative cooling in an adaptive hydrocode in order to investigate the impact of a clumpy, multi-phase interstellar medium.  We find that including cooling results in a very different morphology for the gas in the tail, with a much wider range of temperatures and densities.  The tail is significantly narrower in runs with radiative cooling, in agreement with observed wakes.  In addition, we make detailed predictions of \ion{H}{I}, H$\alpha$ and X-ray emission for the wake, showing that we generally expect detectable \ion{H}{I} and H$\alpha$ signatures, but no observable X-ray emission (at least for our chosen ram-pressure strength and ICM conditions).   We find that the relative strength of the H$\alpha$ diagnostic depends somewhat on our adopted minimum temperature floor (below which we set cooling to zero to mimic physics processes not included in the simulation).

\end{abstract}


\section{Introduction}

Galaxies orbiting within a cluster may undergo a number of interactions that are specific to dense environments.  These include interactions between intracluster gas and the galaxy, such as ram pressure stripping and starvation, interactions between pairs of galaxies, such as harassment, and interactions between the galaxy and the cluster gravitational potential, such as tidal stripping.  The relative importance of these various mechanisms in transforming galaxy morphology, color, and gas content has been the subject of much debate, with large surveys helping to disentangle the various drivers (e.g. Moran et al 2007; van den Bosch et al. 2008).  An alternative way to gauge their relative importance, and to get a deeper understanding of the processes themselves, is to look for observational signs of the different interactions.  This has been attempted recently in the context of ram pressure stripping (Vollmer \& Hutchmeier 2007).  One difficulty with this approach is relating the specific observations to the underlying mechanism: how long does a given signature last and what does it tell us about the physical conditions in the galaxy and cluster?  In this paper, we will use high-resolution numerical simulations of ram pressure stripping (RPS) to better understand the properties and fate of gas stripped during the encounter, and to predict and interpret observational signatures of RPS.  

The ram pressure of the ICM can strip gas from a galaxy moving at typical cluster velocities (Gunn \& Gott 1972), transforming a gas rich spiral to a gas poor S0.  In addition, the galaxy ISM can be compressed, enhancing the star formation rate (Byrd \& Valtonen 1990; Fujita \& Nagashima 1999).  The ICM can also remove the loosely bound reservoir of gas that accretes onto a galaxy, thus slowly starving it of new fuel to form stars (Larson et al. 1980).  In an earlier work (Tonnesen, Bryan \& van Gorkom 2007), we studied cluster galaxies that had formed and evolved in a cosmological simulation, using an adaptive mesh refinement (AMR) code to highly resolve a single cluster.  The simulation included radiative cooling and star formation, and allowed us to predict theoretically how infalling galaxies lose their gas.  We found evidence for both starvation and stripping effects, showing that most galaxies that lost their gas did so without losing stellar mass, an indication that ram pressure stripping was an effective evolutionary mechanism.   

Observationally, a number of studies point to ram pressure stripping as a common environmental interaction.  Spirals in the center of the Virgo cluster have smaller \ion{H}{I} disks than stellar disks, indicating an interaction that does not affect the stellar component of galaxies (Cayatte et al. 1990; Warmels 1988).  Studies of \ion{H}{I} deficiency have shown that galaxies in clusters have less neutral hydrogen than their counterparts in the field (see the review by Haynes et al. 1984).  Solanes et al. (2001) studied \ion{H}{I} deficiency in a sample of 18 cluster regions, and found that \ion{H}{I} deficiency decreases smoothly out to large projected distances from cluster centers. However, the first sightings of head-tail structures in late-type galaxies were radio continuum observations of three galaxies in A1367 (Gavazzi \& Jaffe 1987).  The discovery of these tails, with a maximum length of 30 kpc, was followed up in both \ion{H}{I} and H$\alpha$ (Sullivan et al. 1981; Gavazzi 1989; Dickey \& Gavazzi 1991; Gavazzi et al. 1995).  Observations of displaced \ion{H}{I} within the optical disk in the direction of the radio continuum tails and  \ion{H}{II} regions at the edge of the ``head" of the galaxy supported a ram pressure stripping scenario.  

In a recent, more sensitive and more highly resolved \ion{H}{I} imaging study of Virgo, Chung et al. (2007) found a number of one-sided \ion{H}{I} tails pointing away from the cluster center (tail lengths between 13 and 32 kpc). These galaxies are likely falling in for the first time and gas is already being removed at large projected distances from the cluster center.  In addition, detailed investigations of a few individual galaxies using multiple wavelengths have begun to unravel their probable histories (e.g. Crowl et al. 2005; Chung et al. 2005).  For example, observations of NGC 4522 indicate that the galaxy is undergoing ram pressure stripping, although it is outside of the high density ICM (Kenney, van Gorkom, \& Vollmer 2004).  All of these observations indicate that ram pressure stripping does occur within the cluster environment at a range of distances from the cluster center.  

Recent deep observations of clusters have revealed very long gas tails in \ion{H}{I} (Oosterloo \& van Gorkom 2005; Koopmann et al. 2008, Haynes et al. 2007).  Oosterloo \& van Gorkom find that the $\sim$120 kpc tail of NGC 4388 in the Virgo cluster is well explained by ram pressure stripping.  However, Koopmann et al. (2008) hesitate to attribute the tail of the Virgo pair NGC 4532/DDO 137 to ram pressure stripping because it is more than 1.5 times farther from the cluster center than any other \ion{H}{I} tail in the Virgo cluster, and at 500 kpc is an order of magnitude longer than those found in the simulations of ram pressure stripping by Roediger \& Br\"uggen (2008) and those observed by Chung et al. (2007).  Haynes et al. (2007) also observed a tail that is distant from the cluster center, and used the velocity distribution of the gas in the tail to conclude that the tail associated with NGC 4254 was stripped due to galaxy harassment.  The simulation by Duc \& Bournaud (2008) shows that a single fast flyby could have produced the tail.  

There have also been an increasing number of observations of gas tails in H$\alpha$ (Kenney et al. 2008; Gavazzi et al 2001; Yagi et al. 2007; Yoshida et al. 2004a,b; Yoshida et al 2008; Yoshida et al 2002; Sun, Donahue \& Voit 2007).  The H$\alpha$ tails have filamentary structure and tend to be narrow.  The suspected cause of the H$\alpha$ emission varies from case to case.  For example, Yoshida et al. (2004a) find that most of the emission near NGC 4388 is from stripped gas that has been ionized by the galaxy's AGN.  Gas heating leading to H$\alpha$ emission could also be caused by thermal conduction from the ICM or turbulent shock heating as suggested by both Kenney et al (2008) and Yoshida et al. (2004a,b).  In a different scenario, RB 199 is likely emitting in H$\alpha$ due to star formation: it is probable that this galaxy has undergone a recent merger, making dense, star-forming gas easier to strip (Yoshida et al 2008).   

Tails have also been observed in X-rays,  frequently opposite a sharp edge in galactic X-ray emission or a bow shock (Sun et al. 2006; Kim et al. 2008; Irwin \& Sarazin 1996; Sun \& Vikhlinin 2005; Machacek et al 2006; Wang et al. 2004).  Sun \& Vikhlinin (2005) conclude that the tail they observe is likely galactic gas that has been heated by supernovae explosions and then removed via Kelvin-Helmholtz stripping.  Sun et al. (2006) interpret the tail of the small late-type galaxy ESO 137-001 to be the cool stripped ISM mixed with the hot ICM.  Wang et al. (2004) find a long, 88 kpc, tail from a disturbed disk-like galaxy.  Like Sun et al. (2006), these authors conclude that it is a mixture of cool stripped gas from the galaxy and the ICM, largely because the X-ray emitting gas in the tail is cooler and denser than the surrounding ICM.  Kim et al. (2008) find that the high metallicity of the tail from the large elliptical galaxy NGC 7619 indicates that it is gas from the galaxy, but as it is one of the dominant galaxies in its group, it is possible that the tail is actually sloshing of hot halo gas.  

Whether ram pressure stripping can cause these tails and how they survive in the ICM is unknown.  It is unclear how common long neutral tails are, as Vollmer \& Hutchmeier (2007) examined the surrounding $\sim$60 kpc around 5 \ion{H}{I} deficient galaxies in Virgo, and did not find any excess stripped gas.  Also, the relationship between the gas tails of different temperatures is poorly understood. In fact, three of the long gas tails mentioned above have been observed in multiple wavelengths.  NGC 4388 has a 120 kpc \ion{H}{I} tail (Oosterloo \& van Gorkom 2005), and an extended emission line region consisting of many faint gas filaments emitting in H$\alpha$ close to the galactic disk (Yoshida et al. 2002; Yoshida et al. 2004a).  As discussed above, although this galaxy has an AGN that is heating stripped gas (Yoshida et al. 2004), gas heating leading to H$\alpha$ emission could also be caused by thermal conduction from the ICM or turbulent shock heating.  Another galaxy with multiwavelength emission is ESO 137-001 (Sun et al. 2006; Sun et al. 2007),  with a tail detected in both H$\alpha$ and X-ray.  NGC 4438 also may have a very long tail observed in H$\alpha$ (Kenney et al. 2008), while much smaller \ion{H}{I} and CO tails have been observed (see Vollmer et. al. 2009 and references therein).  Vollmer et al. (2009) found that some of the nearby H$\alpha$ emitting gas is spatially coincident with CO gas, while some is not and is moving at higher velocities.  There are three possible explanations for this difference in velocities:  1) differential stripping (less dense gas is stripped more efficiently from the galactic disk);  2) differential acceleration (less dense gas is accelerated more quickly by the stripping wind); or 3) a combination of differential stripping and differential acceleration.    

Simulations focusing on the tails of stripped galaxies have found somewhat differing results.  Roediger \& Br\"uggen (2008), using an adaptive mesh refinement hydrodynamics code, FLASH, find that for most ram pressures, observable tail lengths are about 40 kpc, with only three of their 31 tail length measurements having lengths over 80 kpc.  However, these simulations did not include radiative cooling.  Kapferer et al. (2009), using a smooth particle hydrodynamics code that includes radiative cooling and star formation, find that tails can survive with cool, dense clumps (T $<$ 2 $\times$ 10$^5$ K) out to well over 100 kpc.  

In our previous work (Tonnesen {\&} Bryan 2009, from now on TB09), we studied a detailed galaxy simulation that included radiative cooling in order to simulate a multiphase ISM.  We found that radiative cooling led to a mixture of high and lower density gas in the disk, and the less dense gas was quickly stripped.  The ICM wind was then able to stream through holes in the disk and ablate the surviving dense clouds.  This is different from the scenario without cooling in which galactic gas is only stripped from the edges of the disk and any gas in the shadow of the galaxy is largely protected from the ICM wind.  Although we focussed on the remaining gas disk in that paper, we noticed that the stripped tail was highly structured.  We found that, although the densest gas clouds, with densities of molecular clouds, were not directly stripped from the galaxy, there was still a lot of structure in the stripped tail, with relatively dense clouds surviving for hundreds of Myr.      

In this paper, we run a set of high resolution simulations (about 40 pc resolution, which is small enough to marginally resolve giant molecular clouds) to understand how a multiphase ISM could affect the survival and structure of ram pressure stripped gas.  It is important to recognize that we do not attempt to include all of the physics involved in the ISM, focusing on how density fluctuations that are observed in the multiphase ISM of galaxies can affect gas tails.  Following Roediger \& Br\"uggen (2008), we focus on the stripped tails of gas, specifically examining the length, width, and substructure of the tails (see also Roediger, B\"ruggen \& Hoeft 2006).  Including radiative cooling allows us to estimate the density of the gas, and emission from \ion{H}{I}, H$\alpha$, and X-ray gas separately.  The questions that we highlight are:  can cool gas tails survive in the ICM with turbulence and shock heating of the stripped gas? How can bright H$\alpha$ and X-ray emission be induced in a tail of gas?  Finally, to what velocities is stripped gas accelerated, and does differential stripping occur in gas disks?

The paper is structured as follows.  After a brief introduction to our methodology, we provide the general characteristics of our simulation (\S 2.1-3).  We introduce the parameters of our specific simulations in \S 2.4.  We then (\S 3) discuss our results, specifically focusing on the density, temperature, and velocity structure in the gas tail.  In \S 4 we compare our results to observations.  We discuss our choice of radiative cooling floor and resolution in \S 5.1-2, and the impact of additional physics to our results in \S 5.3-5.  Finally, we conclude in \S 6 with a summary of our results and predictions for observers.

\section{Methodology}

We use the adaptive mesh refinement (AMR) code {\it Enzo}.   To follow the gas stripped from our simulated galaxy, we employ an adaptive mesh for solving the fluid equations including gravity (Bryan 1999; Norman \& Bryan 1999; O'Shea et al. 2004).  The code begins with a fixed set of static grids and automatically adds refined grids as required in order to resolve important features in the flow.

Our simulated region is 311 kpc on a side with a root grid resolution of $128^3$ cells.   We allow an additional 6 levels of refinement for our runs that include radiative cooling, for a smallest cell size of 38 pc.  We refine our simulation based on the local gas mass, such that a cell was flagged for refinement whenever it contained more than 4892 $\msun$ (our standard, best mass resolution), or 41745 $\msun$ (our lower mass resolution, used in TB09).  We found that both parameters refined most of the galactic disk to 38 pc resolution by the time the wind hit, although the disk fragmentation was better resolved in the higher resolution run, resulting in smaller mass clumps.  The high mass resolution run also refined much of the wake to a spatial resolution of about 76 kpc, while the run with lower mass resolution generally resulted in a factor of 2-4 lower spatial resolution in the wake.

The simulation includes radiative cooling using the Sarazin \& White (1987) cooling curve extended to low temperatures as described in Tasker \& Bryan (2006).  To mimic effects that we do not model directly (such as turbulence on scales below the grid scale, UV heating, magnetic field support, or cosmic rays), we cut off the cooling curve at a minimum temperature $T_{\rm min}$ so that the cooling rate is zero below this temperature.  In the simulations described here we use either $\tmin = 8000$ K, or $\tmin = 300$ K.  Both allow gas to cool below the threshold for neutral H formation.  In TB09, we found that the minimum temperature affected the range of masses and sizes of clouds forming in the disk.  This, in turn, resulted in changes in the timescale for gas loss from the disk, although the total amount of gas lost was similar.  In this paper, as we will show in more detail later, we find that the structure of the wake also depends somewhat on $\tmin$.

\subsection{The Galaxy}

Our galaxy is placed at a position corresponding to (155.5,155.5,68.42) kpc from the corner of our cubical 311 kpc computational volume, so that we can follow the stripped gas for more than 200 kpc.  The galaxy remains stationary throughout the runs.  The ICM wind flows along the z-axis in the positive direction, with the lower z boundary set for inflow and upper z boundary set as outflow.    The x and y boundaries are set to outflow in all three of our high resolution cases (8000K, 300K, and no cooling) and reflecting in the lower resolution cases.  The different boundary conditions in the two runs had only a very minimal impact, since the box is so wide (311 kpc wide).

We chose to model a massive spiral galaxy with a flat rotation curve of 200 km $s^{-1}$.  It consists of a gas disk that is followed using the adaptive mesh refinement algorithm (including self-gravity of the gas), as well as the static potentials of the stellar disk, stellar bulge, and dark matter halo.  We directly follow Roediger \& Br\"uggen (2006) in our modeling of the stellar and dark matter potential and gas disk.  In particular, we model the stellar disk using a Plummer-Kuzmin disk (see Miyamoto \& Nagai 1975), the stellar bulge using a spherical Hernquist profile (Hernquist 1993), and the dark matter halo using the spherical model of Burkert (1995).  This dark matter halo model is compatible with observed rotation curves (Burkert 1995; Trachternach et al. 2008).  The equation for the analytic potential is in Mori \& Burkert (2000).  

\begin{table}
\begin{center}
\caption{Galaxy Stellar and Dark Matter Constants\label{tbl-const}}
\begin{tabular}{c | c}
\tableline
Variable & Value\\
\tableline
M$_*$ & $1 \times 10^{11}$ M$_{\odot}$ \\
a$_*$ & 4 kpc \\
b$_*$ & 0.25 kpc\\
M$_{bulge}$ & $1 \times 10^{10}$ M$_{\odot}$ \\
r$_{bulge}$ & 0.4 kpc \\
r$_{DM}$ & 23 kpc \\
$\rho_{DM}$ & $3.8 \times 10^{-25}$ g cm$^{-3}$ \\
\tableline
\end{tabular}
\end{center}
\end{table}

The gas is described as a softened exponential disk:  
\begin{equation}
\rho(R,z) = \frac{M_{\rm gas}} {2\pi a^2_{\rm gas}b_{\rm gas}}0.5^2 
{\rm sech} \left( \frac{R}{a_{\rm gas}}\right)
{\rm sech} \left( \frac{|z|}{b_{\rm gas}}\right)
\end{equation}
Given this gas density distribution in the disk, the initial gas temperature and pressure are calculated to maintain the disk in hydrostatic equilibrium with the surrounding ICM in the z direction.  The gas disk's rotational velocity is set so that the combination of the centrifugal force and the pressure gradient of the disk balances the radial gravitational force.  We taper the gas disk smoothly by multiplying the gas density distribution by $0.5(1 + \rm cos(\pi (R - 20$ kpc$)/26$ kpc$))$ for 20 kpc $<$ R $\leq$ 26 kpc.  See our galaxy parameters in Tables \ref{tbl-const} and \ref{tbl-gconst}.

\begin{table}
\begin{center}
\caption{Gas Disk Constants\label{tbl-gconst}}
\begin{tabular}{c | c}
\tableline
M$_{gas}$ & $1 \times 10^{10}$ M$_{\odot}$ \\
a$_{gas}$ & 7 kpc \\
b$_{gas}$ & 0.4 kpc\\
\tableline
\end{tabular}
\end{center}
\end{table}

To identify gas that has been stripped from the galaxy we also follow a passive tracer which is initially set to 1.0 inside the galaxy and $10^{-10}$ outside.  In the following analysis, we will use a minimum tracer fraction of 25\% to find gas stripped from the galaxy (our conclusions do not change when we use 10\%).

\subsection{ICM Conditions}\label{sec:ICM}

The galaxy initially evolves in a static, high-pressure ICM with $\rho=$ 9.152 $\times$ 10$^{-29}$ g cm$^{-3}$ and $T = $ 4.15 $\times$ 10$^6$ K, to allow cool, dense gas to form in the galaxy.  This naturally generates a multiphase ISM (see Tasker \& Bryan 2006 and TB09 for more discussion of the ISM properties).  

After 155 Myrs, we reset the boundary conditions to generate a constant ICM inflow along the inner z-axis, which is always face-on to the galaxy.   We choose to study the highest of the three ram pressure strengths examined in TB09 (this case results in the largest mass in the wake).  We choose $P_{\rm ram} = \rho v^2_{\rm ICM} = 6.4 \times 10^{-12}$ dynes cm$^{-2}$.  For the corresponding ICM physical conditions, $\rho$, $T$, and $v_{\rm ICM}$, we use the results from our earlier cluster simulation to find the mean density, velocity, and temperature encountered by infalling galaxies for a typical cluster mass, given the ram pressures value (Tonnesen, Bryan \& van Gorkom 2007).  See TB09 for other details regarding the numerical setup (we use case PHRCW in TB09).  Because we are interested primarily in the impact of radiative cooling on a gas stripped tail, we only consider a case with a face-on wind, again because this will result in the most gas in the tail.

\subsection{Suite of Simulations}

We run two main simulations of a ram-pressure stripped galaxy including radiative cooling using a temperature-dependant cooling curve from Sarazin \& White (1987) extended to either $\tmin = 8000$ K or $\tmin = 300$ K, as described earlier.  In addition, we run a simulation without radiative cooling.  In this way, we can investigate the impact of cooling to various temperature floors.  These three runs all have the highest mass resolution and provide the main results of this paper.

All simulations have the same maximum refinement level of $l=6$ (i.e. the same best spatial resolution of 38 kpc), but we can compare both of our radiatively cooled runs to ones with a factor of 8 lower mass resolution  (as described earlier).  This allows us to investigate the effect of numerical resolution.

A final comparison simulation has the same refinement criteria as our highest mass resolution runs, but calculates cooling rates by following the non-equilibrium ionization fractions of hydrogen and helium, and directly computing their cooling rates.  We do not include metals in this cooling calculation, which leads to slower cooling of hot ISM gas.  This in turn results in a stripping rate which lies between our radiative cooling runs and a run without radiative cooling (largely because of the smaller amount of fragmentation in the disk).  This comparison run was primarily done to allow us to check our simple calculation of H$\alpha$ emission -- described in the next section -- by using the hydrogen ionization state as computed by the code.

\subsection{Projections}\label{sec:projection}

Enzo outputs the density and temperature of the gas in each cell.  To transform these values into \ion{H}{I} column density and H$\alpha$ intensity, we used Cloudy, version 08.00 of the code last described by Ferland et al. (1998).  Using a grid of temperatures and densities, we calculated the hydrogen neutral fraction and H$\alpha$ emissivity.   In the Cloudy calculation, we included CMB radiation, the cosmic ray background, bremsstrahlung radiation from the ICM and the 2005 version of the Haardt \& Madau (2001) $z=0$ metagalactic continuum, as implemented by Cloudy.  We found that including the local interstellar radiation field emission resulted in lower amounts of neutral gas.  Since much of our gas is very distant from the galaxy, we decided not to include this radiation.  We also found that removing bremsstrahlung radiation did not change any of the values we considered. 

We chose to calculate the neutral fraction and H$\alpha$ emissivity for a thin plane-parallel gas cloud of width 100 pc.  We selected this width because it loosely corresponds to the cell size of most of the gas in the highly resolved tails, and accounts approximately for radiative transfer effects.  If we assumed the radiative thin limit, it would decrease the amount of \ion{H}{I} we predict, and increase the H$\alpha$ emission for dense, low-temperature gas.  Ideally, we would include the radiation field with radiative transfer directly in the simulation, but this is not yet feasible (and only has a mild impact on the dynamics); instead we post-process these results to get reasonable predictions for the ionization fraction and H$\alpha$ emissivity (see Furlanetto et al. 2005 for a discussion of various approaches in the context of Ly$\alpha$ emission).  We do compare our projections with ones made assuming clouds sizes of 10 pc and 1 kpc.   
 
To create X-ray surface brightness projections, we use a spectral lookup table for low density gas with a metallicity of 0.3 solar computed using a Raymond-Smith code (Raymond \& Smith 1977), as updated in XSPEC (Arnaud 1996).  The X-ray band we use is 0.5 keV to 2.0 keV, following Sun et al. (2006).  

\begin{figure*}
\begin{center}
\includegraphics[scale=0.8,trim= 30mm 0mm 26mm 0mm,clip]{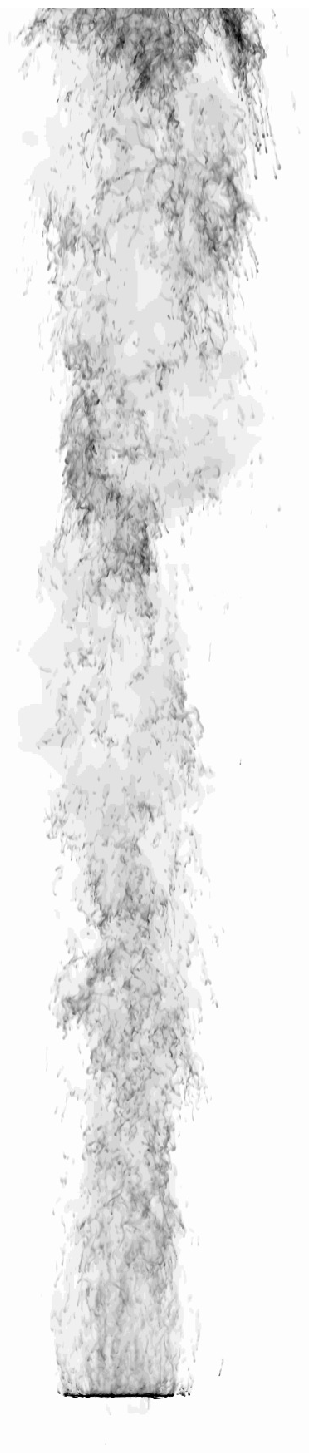}
\includegraphics[scale=0.8,trim= 12mm 0mm 12mm 0mm,clip]{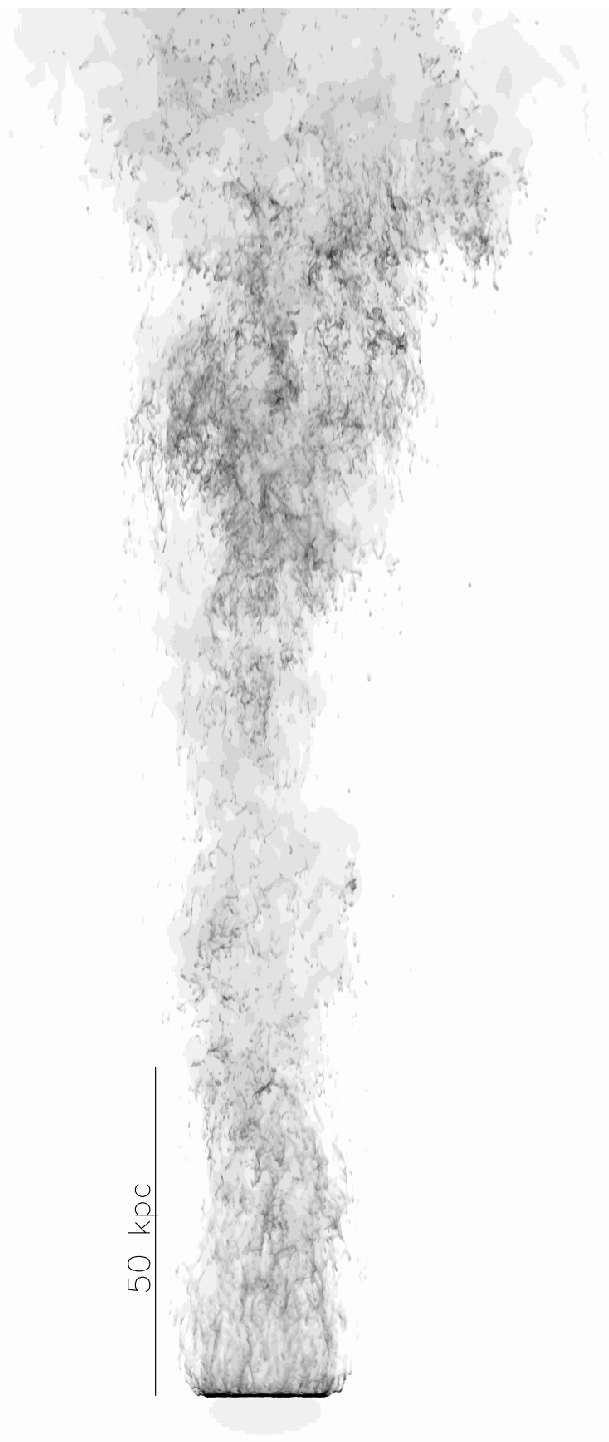}
\includegraphics[scale=0.8,trim= 2mm 0mm 2mm 0mm,clip]{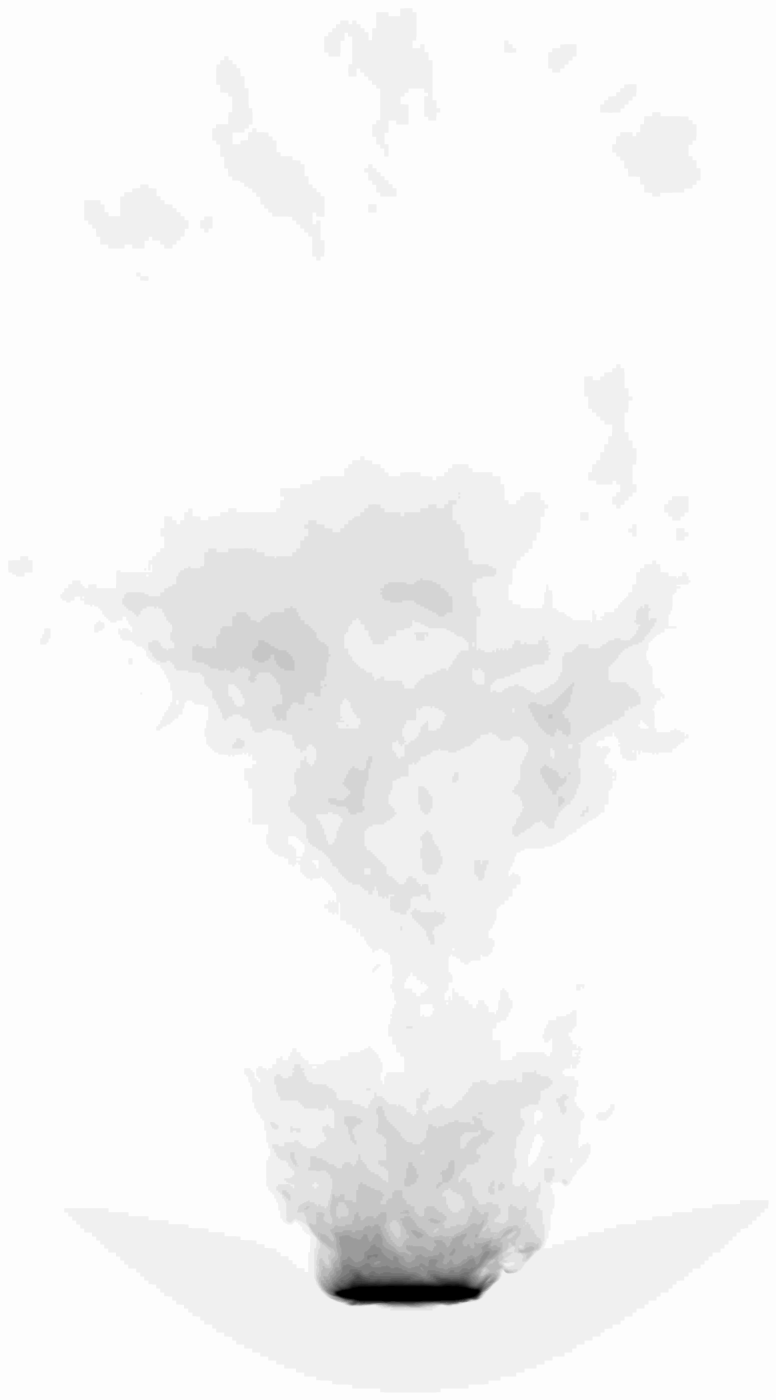}
\includegraphics[scale=0.6]{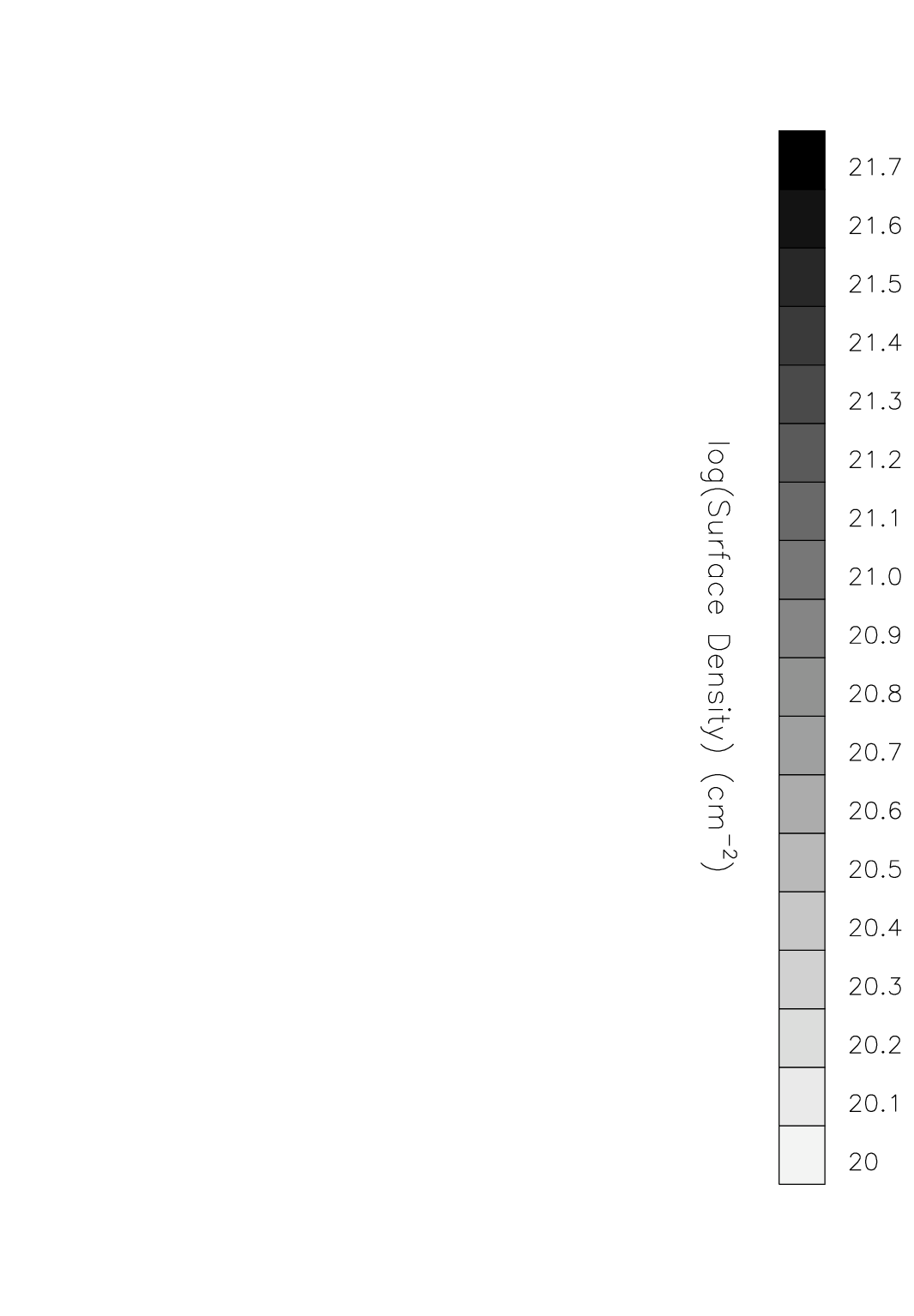}
\end{center}
\caption{Surface density of all gas 500 Myr after the wind has hit the galaxy for our runs with radiative cooling to $\tmin = 300$ K (left), $\tmin = 8000$ K (center) and for the comparison run with no radiative cooling (right).  The images are shown with a logarithmic stretch.  Note the large amount of structure, the length, and the lack of flaring in the tail with radiative cooling, compared to the tail without radiative cooling.  The radiatively cooled tails are cut off in this projection 212 kpc from the galaxy.}
\label{fig:gas}
\end{figure*}

\section{Tail Attributes}

In this section, we examine the physical characteristics of the wakes, focusing on the wake morphology, phase diagrams and the velocity of the stripped gas.  Then, in the following section, we turn to observational diagnostics.

\begin{figure*}
\includegraphics{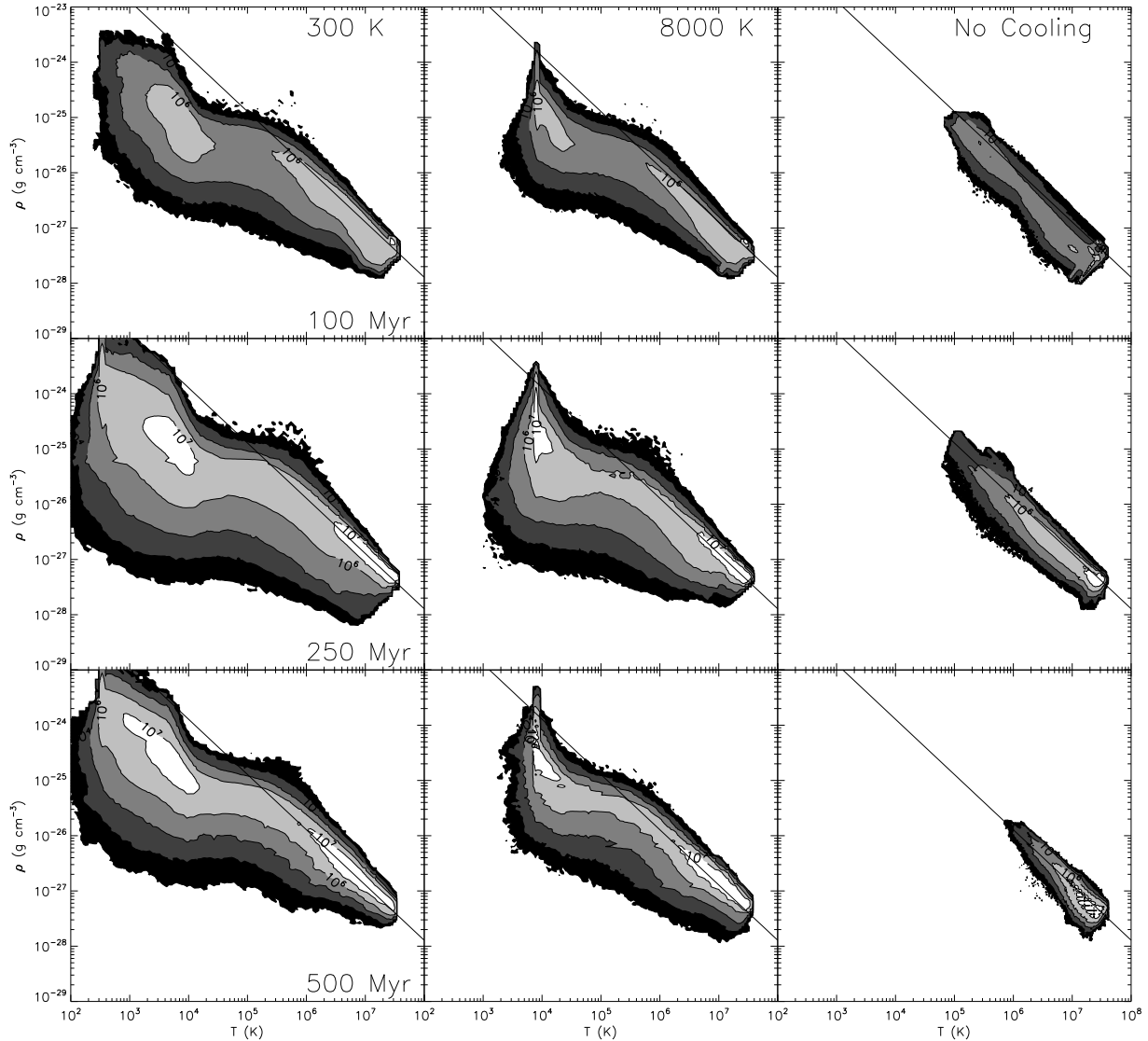}
\caption{Contour plots showing the mass in gas at different densities and temperatures for our $\tmin = 300$ K run on the left, $\tmin = 8000$ K in the middle, and the comparison run without radiative cooling on the right.  The contours are log mass. For a gas cell to be included in the contour plot, it must have at least 25\% of its mass originating from the galaxy.  The line denotes the surrounding ICM pressure.  Note that at lower temperatures the gas is below the ICM pressure due to rapid cooling (see text).  Note also that the range of pressures in the simulations with cooling are much larger than the comparison run without.  See section \ref{sec:rhot} for discussion.}
\label{fig:rhot}
\end{figure*}

\subsection{Morphology}

We first consider the morphological characteristics of our simulations.  See Figure \ref{fig:gas} to compare the surface density of the $\tmin = 300$ K (left), $\tmin = 8000$ K (center) simulations and the run without radiative cooling (right) at a time 500 Myr after the wind has hit the galaxies.  Clearly there are important differences in the morphology of the resulting tails, although, remarkably, the amount of mass lost by this time differs by only 20\% between $\tmin = 8000$ K and the run without radiative cooling.

We find there are three primary changes that radiative cooling makes in the tail morphology.  First, because of the formation of clouds in the disk, the runs with cooling show more structure in the tail.  As we will see below, there is a wide range of densities and temperatures, with lots of small pressure-confined clouds.  This is entirely missing in the no-cooling run and makes a crucial difference in some of the observational diagnostics.  This result is largely independent of $\tmin$, as long as cooling can generate a clumpy ISM.

Second, the tails in the cooling runs are more extended at a given time, reaching more than 200 kpc past the galaxy within 500 Myr of the wind hitting the galaxy.  In the run without radiative cooling, some gas has gone as far as 150 kpc, but the majority of the tail is closer than 125 kpc from the galactic disk.   In all of our simulations, stripped gas begins moving away from the galaxy as a group, however, in the cases with radiative cooling, with time the material is stretched out as the velocities vary (we address the velocity distribution in more detail later).

Finally, the stripped gas in the radiative cooling runs expands only slightly in the transverse direction as it moves with the ICM wind: in the $\tmin = 8000$ K run, 200 kpc from the disk the tail width is about 80 kpc (500 Myr after the wind has hit the galaxy).  The $\tmin = 300$ K run is even narrower, with a tail width of 40 kpc 200 kpc downstream of the disk.  This is much less flaring than seen in in our no-cooling run: 500 Myr after the wind has hit the galaxy, the tail is 85 kpc wide at distance 125 kpc downstream from the disk.  Although this is total gas surface density, which cannot be observed, it is notable that most observed tails are long and narrow, and do not seem to flare (for example, see the images in Chung et al. 2007).   

The no-cooling results are in qualitative agreement with the simulations of Roediger \& Br\"uggen (2008), who also find the tails to have a decreasing density profile with distance from the galaxy.  We find that including radiative cooling permits higher density and lower gas to form in the disk and hence to be stripped, leading to high density gas throughout the tail.  Our surface density projection grossly resembles that of Kapferer et al. (2009), who also include radiative cooling in their simulations, in that we both have elongated gas clumps in the wind out to large distances above the galactic disk.  However, at less than a kiloparsec in diameter, most of our clumps are much smaller (about 10\% in linear size) and much less massive than the clumps in the Kapferer et al. (2009) simulation.

\subsection{Gas Temperature and Density}\label{sec:rhot}

In Figure~\ref{fig:rhot}, we show the mass-weighted distribution of density and temperature for gas in the wake that originated from the galaxy.  The plots include all of the gas located between 10 kpc and 240 kpc above the disk that has at least 25\% of the gas originating in the galaxy.  We show these plots for three different times after the wind has hit the galaxy: 100 Myr, 250 Myr, and 500 Myr for the runs with radiative cooling to 300 K (left), with radiative cooling to 8000 K (middle) and without cooling (right).  The solid line shows the background ICM pressure of 1.76 $\times$ 10$^{-12}$ dynes cm$^{-2}$.  

Once again, we see a clear difference between the runs with and without radiative cooling.  To begin, we concentrate on a comparison between the no-cooling run and the simulation with cooling to $\tmin = 8000$ K.  We explore three distinct changes in the phase diagram.  

\begin{figure}
\includegraphics[scale=6.5,trim= 37.5mm 20mm 32mm 127mm,clip]{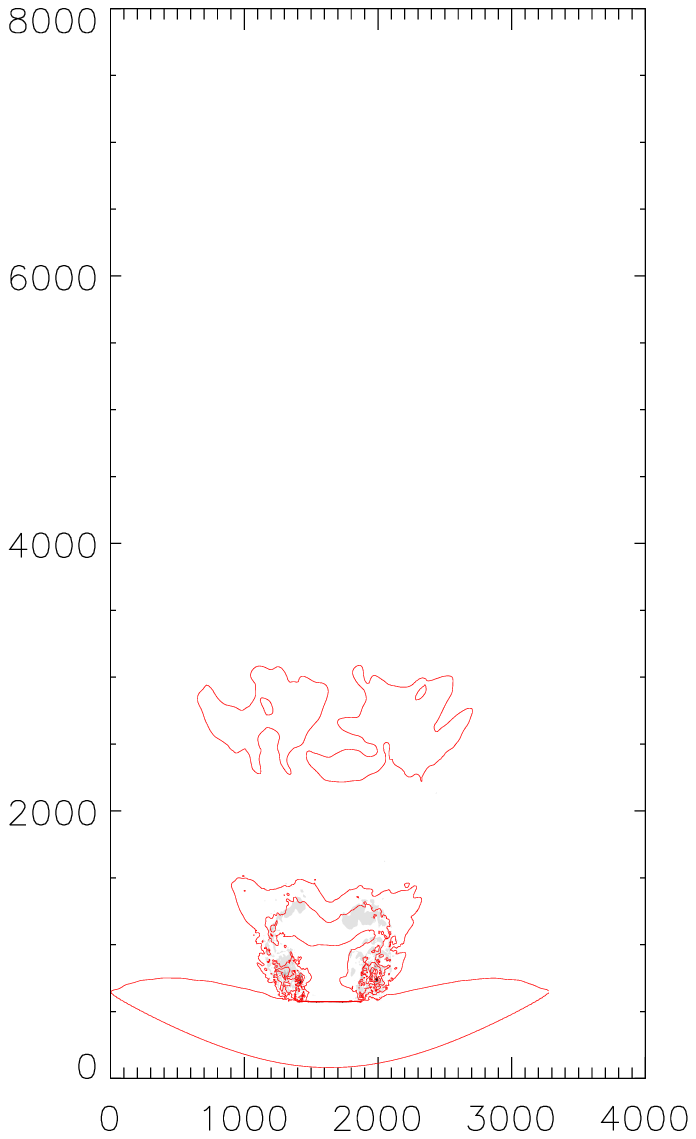}
\caption{Zoom-in of a region of the wake from the run with cooling to 8000 K near the galactic plane 180 Myr after the wind has hit the disk.  The greyscale shows temperature (lowest temperature is black), while the red contours indicate pressure.  The outermost contours have the largest pressure.  The slice is 6 kpc in depth, and 20 kpc $\times$ 37 kpc in size.  Note that the lowest pressure contours surround the coolest gas clouds (in black).  See section \ref{sec:rhot} for a discussion.}
\label{fig:rhot2}
\end{figure}

First, there is the simple fact that gas exists at much higher density and lower temperature in the runs with cooling, even at constant pressure.  This is mostly due to the formation of dense clouds in the galaxy ISM before stripping (see also TB09), but it is clear that cooling plays a role even at late times: in the no-cooling run, the distribution becomes narrower as time goes on, as the gas is heated and mixed, while, if anything, the cooling distributions become more extended.  This higher-density gas shows up clearly in the tail morphology as shown in Figure~\ref{fig:gas}.

Second, the cooling runs show a wider spread of density and pressure at fixed temperature; this is true even at large temperatures, but appears to be most noticeable at lower temperatures.   Turbulence can cause a spread in pressure throughout the gas in the wind, as can be seen both in the run without radiative cooling, and at higher temperatures in our radiatively cooled runs, where cooling is inefficient.  

The final distinctive difference is a clear break in the distribution at $T \sim 10^5$ K, with gas below this temperature falling off the constant pressure line.  This corresponds to the peak of the cooling curve, when the cooling time becomes very short.  In cooler clumps of stripped gas there is an interplay between cooling, which lowers the pressure of the gas, and adiabatic compression or shock heating from the surrounding ICM.  At $T \sim 10^5$ K, radiative cooling decreases the temperature more quickly than compression can increase the pressure, resulting in most of the cool gas lying below the pressure of the ICM.  The peak in the gas pressure in the $\tmin = 8000$ K run occurs where radiative cooling is turned off, which suddenly allows the gas to be recompressed to the ICM pressure.  

To investigate this process in more detail, we look closely at a slice from our $\tmin = 8000$ K simulation in Figure \ref{fig:rhot2}, in which grayscale contours of temperature (with black being the lowest temperatures) are plotted with red contours of pressure, with the outermost contours denoting the highest pressure.  This 6 kpc thick slice is 20 kpc $\times$ 37 kpc, and is taken 180 Myr after the wind has hit the galaxy.  As expected from the above argument, the lowest temperature regions are also the lower pressure regions, indicating that cooling lowers pressure faster than adiabatic compression can increase it.  

A simple timescale argument shows that the cooling and compression times are similar.  For example, at $T = 10^5$ K, the cooling time for a clump density of 0.1 cm$^{-3}$ is about the same as the sound crossing time across the clump, given a size of 1 kpc.  As we find that many of our clouds are smaller than this size, compression must be acting more slowly than estimated using this simple calculation.  

The $\tmin = 300$ K run is similar but shows that gas can continue to cool below 8000 K in dense regions.  The return to high pressures can be seen at this lower $\tmin$, otherwise the gas is in a rough pressure equilibrium below $T \sim 10^4$ K, but at a pressure approximately an order of magnitude below the ICM pressure.  

\begin{figure*}
\includegraphics{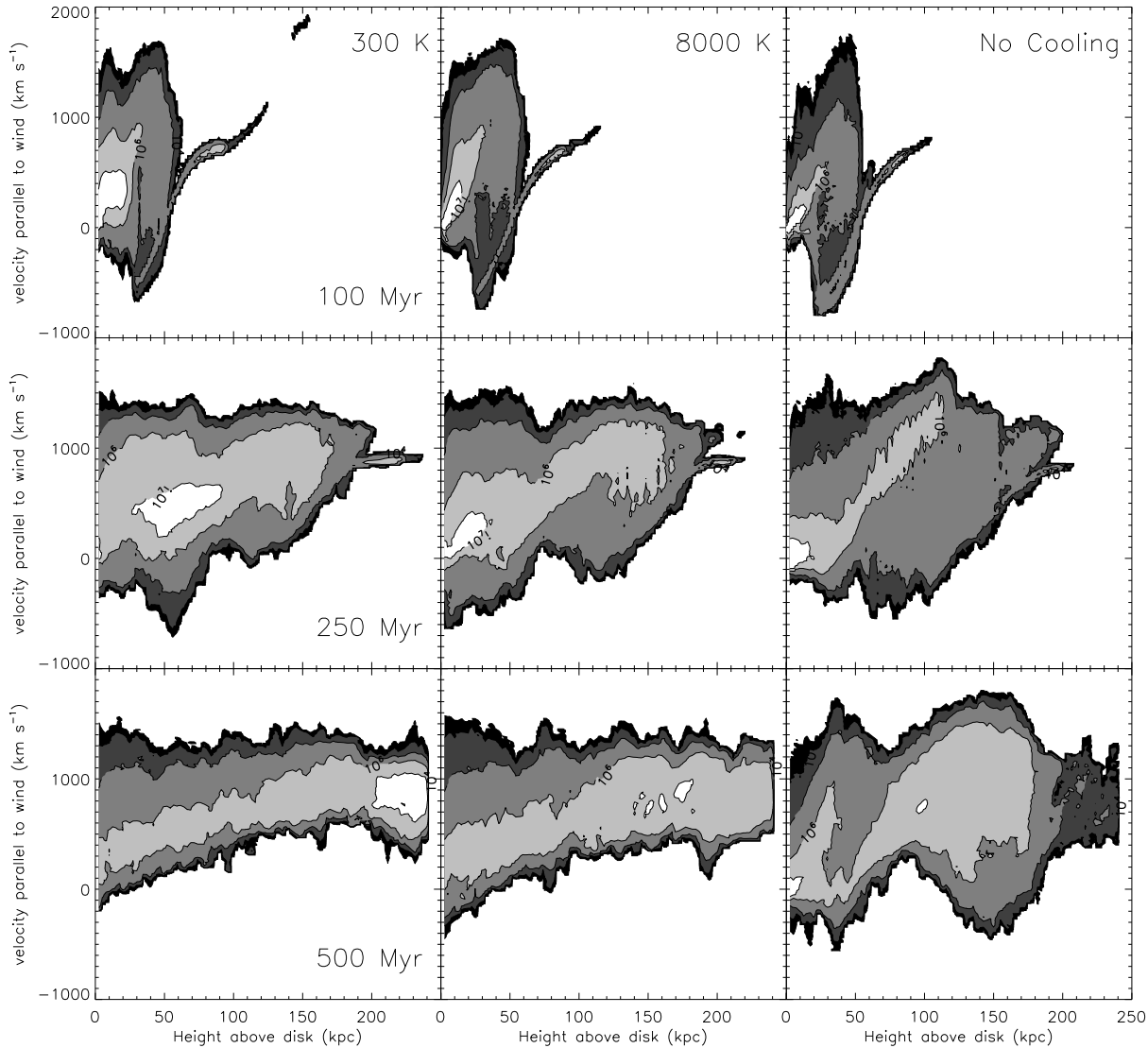}
\caption{Contour plots of gas mass at different velocities parallel to the wind (perpendicular to the disk) for the $\tmin = 300$ K, $\tmin = 8000$ K, and no cooling cases (going left to right).  The contours are plotted as a function of height above the disk.  Again, only gas cells with at least 25\% of their mass originating from the galaxy are counted.  Very little of the disk gas is accelerated to the ICM wind velocity of 1413 km/s.  See section \ref{sec:vel} for a discussion.}
\label{fig:velocities}
\end{figure*}

\subsection{Velocity Distribution}\label{sec:vel}

In this section we examine the velocity structure of the stripped gas in our three main runs, from left to right:   $\tmin = 300$ K, $\tmin = 8000$ K, and the no-cooling run.   Figure~\ref{fig:velocities} shows the gas z-velocity, parallel to the ICM wind at times 100 Myr, 250 Myr, and 500 Myr after the wind has hit the galaxy.  Again we are only considering gas that originated from the disk (a 25\% minimum tracer fraction).  We plot contours of gas mass as a function of velocity and distance above the disk.  In our runs with radiative cooling, we find that the majority of stripped gas is never accelerated to the ICM wind speed of 1413 km/s.  In fact, 500 Myr after the wind has hit the galaxy, the main contour of gas asymptotes at about 900 km/s from about 150 kpc above the disk in both cooling runs. 

Our comparison run without radiative cooling evolves very differently.  All three cases look similar in velocity space 100 Myr after the wind has hit the galaxy, but by 250 Myr -- in the no cooling case -- the main mass contour of stripped gas is accelerated to very close to the ICM speed (1413 km/s) at a height 100 kpc above the disk, with a narrow velocity width of only $\sigma \sim$ 250 km/s.  The radiative cooling cases accelerate more slowly with distance above the disk, with none of the main contour of stripped gas at the ICM velocity at 250 Myr.  By 500 Myr after the wind has hit the disk, the radiative cooling cases show a narrow range in their tail velocity distributions, but the case without cooling has a very wide range of velocities, with stripped gas moving at both the ICM velocity and falling back towards the disk (at a height of 150 kpc above the galaxy).  The width of the wake in the velocity distribution is a measure of the strength of turbulence seen in the wake.  At late times, the radiatively cooled cases have a characteristic width of $\sigma \sim$ 700 km/s, while the no-cooling case seems to be more turbulent, with a characteristic width of $\sigma \sim$ 1000 km/s.

To understand this difference, we turn back to the different morphology of the tails.  The important difference between our cooling and no-cooling runs is that radiative cooling forms much more structure in the tail.  These dense clumps may be more difficult to accelerate and less affected by turbulence, so will not shift laterally into the shadow of the disk.   In the no-cooling run, the large variation in the wind velocity has to do with the large eddies that form from the interaction between the stripped gas and ICM, as seen in Roediger \& Br\"uggen (2007) (their Figure 6).  These eddies move the gas into the shadow of the disk, where the gas can fall back onto the galaxy.  Also, both this work and Roediger \& Br\"uggen (2008) find that some stripped gas is accelerated to the ICM velocity about 100 kpc downwind of the galaxy (this is true even though Roediger \& Bruggen (2008) vary their wind velocity while ours remains constant at 1413 km/s).  

\begin{figure*}
\includegraphics[scale=1.05, trim=0mm 35mm 0mm 65mm, clip]{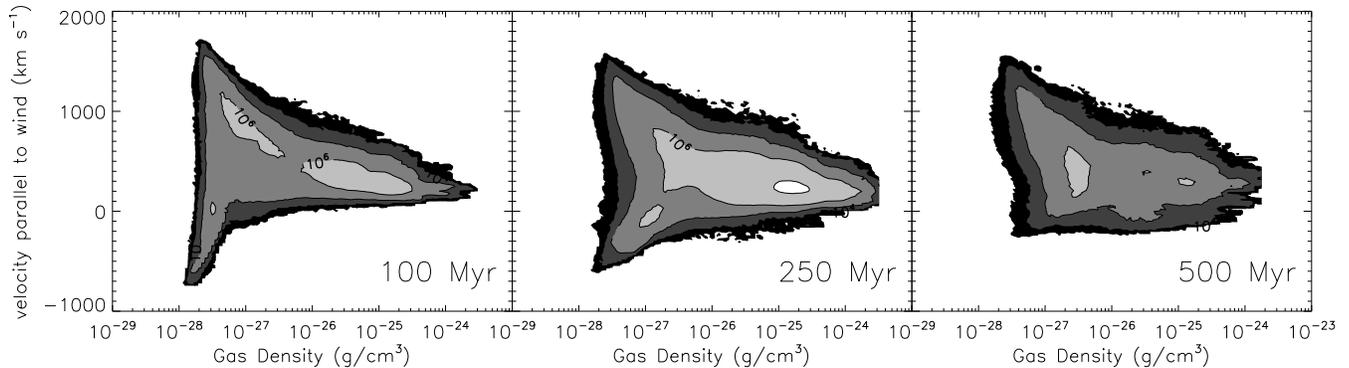}
\caption{Contour plots of gas mass at different velocities parallel to the wind (or perpendicular to the disk) for the $\tmin = 8000$ K case.  The contours are plotted as a function of gas density.  Again, only gas cells with at least 25\% of their mass originating from the galaxy are counted, and we only include gas between 2 and 50 kpc above the disk.  We can see differential stripping and acceleration.  See section \ref{sec:vel} for discussion.}
\label{fig:vrho}
\end{figure*}

In addition to looking at the velocity structure of all stripped gas, we can see how different density gas is affected by the ICM wind, looking for differential stripping and acceleration.  In Figure \ref{fig:vrho}, we consider in detail how the velocity in the wind direction is related to gas density.  The gas in these contours is between 2 kpc and 50 kpc above the disk.  The lower density gas is stripped early to a higher velocity, as shown in the top right, while higher density gas has almost no velocity parallel to the wind because the dense clouds are harder to strip and accelerate.  This systematic correlation between density and velocity (differential stripping) has been observed in NGC 4438 with multiwavelength observations (Vollmer et al. 2009).  There is a similar effect in the no cooling case (not shown), but it has a somewhat different origin: it is caused by the exponentially declining density distribution in the disk.  The gas density decreases towards the edge of the disk, which is the only part of the disk stripped by the wind.  

Later, as shown in the right panel of Figure~\ref{fig:vrho}, even higher density gas is eventually accelerated by the wind, although it does not reach the velocities of the lower density gas.  This gradual acceleration of denser material with time is not seen in the no cooling run, again because only the lower density gas at the edges is ever stripped.  The differential velocity as a function of gas density is seen throughout the tail in all of our runs.   It is also notable that negative velocities, denoting gas falling back onto the disk, are more often seen in lower density gas in all of our runs.  In the no cooling run, this is because there is only low density gas in the tail.  In the cooling cases, this may be because only the lower density gas can be moved into the shadow of the disk by turbulence.  This result differs from Kapferer et al. (2009), who find that dense \ion{H}{I} gas falls back onto the disk.  Kapferer et al. (2009) are not the only ones to predict gas re-accretion (e.g. Vollmer et al. 2001; Roediger \& Br\"uggen 2007), although it has not been clearly observed.

We can now compare our simulations to observations.  The stripped gas moving more slowly than the ICM wind, at 900 km/s, is marginally consistent with (although somewhat larger than) the observations of Oosterloo \& van Gorkom (2005), who find the gas from the galaxy is only accelerated by 550 km/s, not the 1500 km/s the authors estimate as the galaxy velocity.  However, if we consider the gas velocity 120 kpc above the disk, it is closer to 700 km/s.  Also, recall that the contours in Figure~\ref{fig:velocities} include gas at all densities.  From Figure~\ref{fig:vrho}, we see that higher density gas moves more slowly than lower density gas.  In fact, at any height above the disk, gas with densities greater than about 0.1 cm$^{-3}$ moves more slowly than 900 km/s, and is frequently centered at about 500 km/s.  This puts our results into good agreement with the observations of Oosterloo \& van Gorkom (2005).

\section{Comparison to Observations}
\subsection{\ion{H}{I}}\label{sec:hi}

\begin{figure*}
\begin{center}
\includegraphics[scale=0.7,trim=10mm 10mm 10mm 0mm, clip]{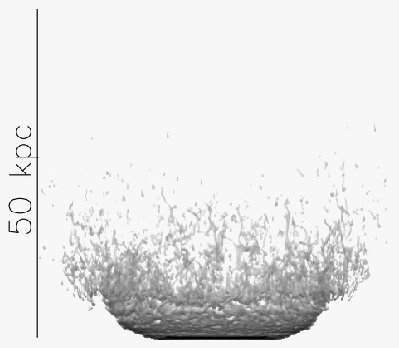}
\includegraphics[scale=0.7,trim=10mm 10mm 10mm 0mm, clip]{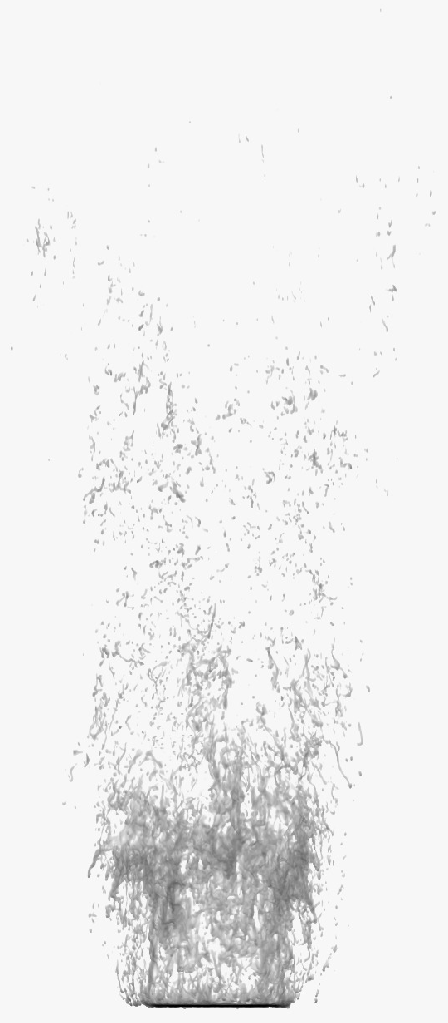}
\includegraphics[scale=0.7,trim=10mm 10mm 10mm 0mm, clip]{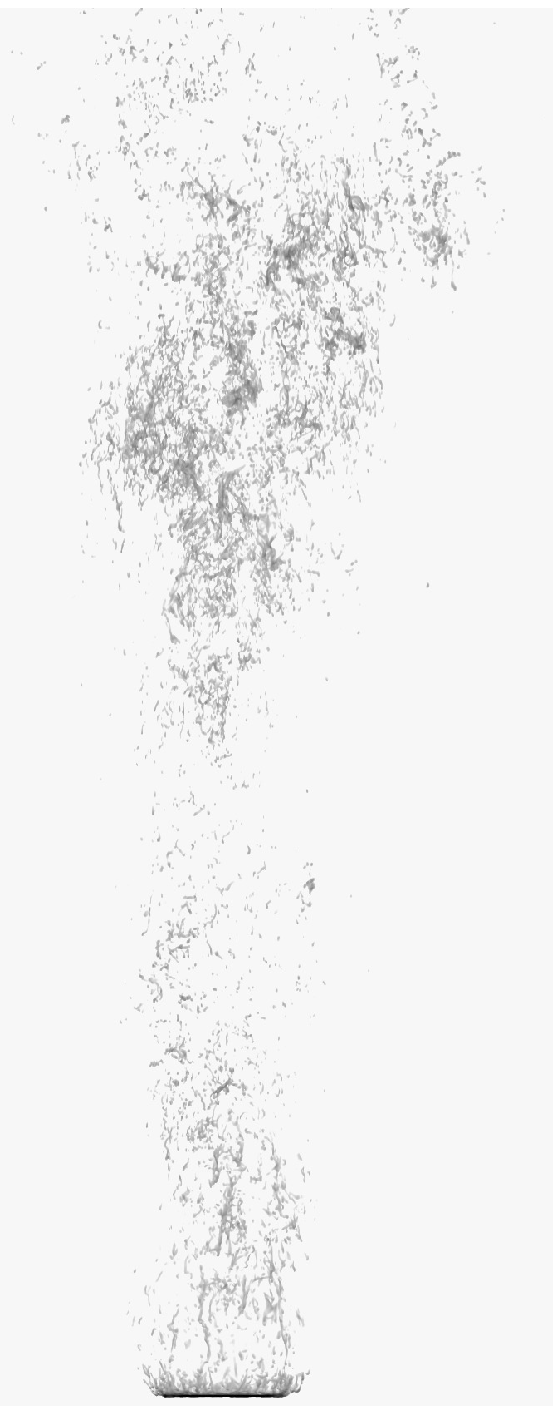}
\includegraphics[scale=0.7]{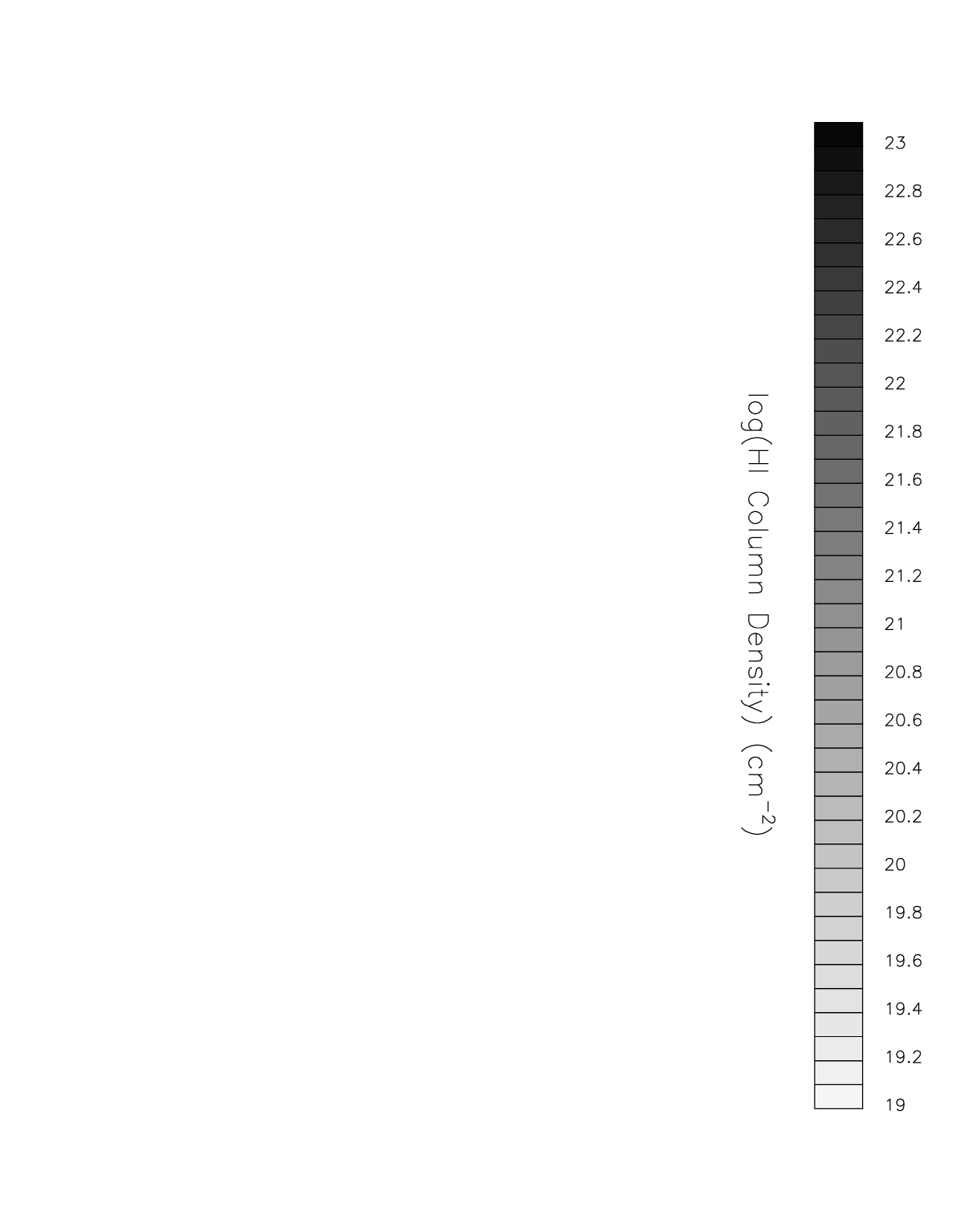}\\
\end{center}
\caption{The column density of \ion{H}{I} gas in the $\tmin = 8000$ K case shown at the 38 pc resolution of our simulation 100 Myr (left), 250 Myr (center), and 500 Myr (right) after the wind has hit the galaxy.  The greyscale shows the log of the column density in units of cm$^{-2}$.  The longest tail is again cut off at 212 kpc.  See Section \ref{sec:hi} for discussion. }
\label{fig:hi}
\end{figure*}

In this section we consider \ion{H}{I} column density, comparing projections of our simulations to observations.  We use recent deep observations (Chung et al. 2007; Oosterloo \& van Gorkom 2005) to select the minimum observable column density of \ion{H}{I} to be 10$^{19}$ cm$^{-2}$.  As discussed in Section \ref{sec:projection}, we use Cloudy to determine the neutral fraction given a temperature and density and then apply that value in our projection routine.  At the resolution of our simulation, we find gas with high column densities at very large distances from the galaxy, and for a long time after stripping begins (see Figure \ref{fig:hi}).  The individual \ion{H}{I} clouds frequently have a head-tail structure, indicating Kelvin-Helmholtz stripping, and indeed the gas clumps are largely shredded in the tail, losing their coherency.  This differs from Kapferer et al. (2009), who found that Kelvin-Helmholtz stripping does not affect the cool clouds in their stripped tails (note that Agertz et al. (2007) have recently argued that SPH has difficulty following the Kelvin-Helmholtz and Rayleight-Taylor instabilities).  There is a gap between the galaxy and region with the most \ion{H}{I} gas at later times.  Oosterloo \& van Gorkom (2005) also observe a gap in \ion{H}{I} column density in the long tail from NGC 4388.

\begin{figure*}
\begin{center}
\includegraphics[scale=0.8, trim= 17mm 0mm 13mm 0mm, clip]{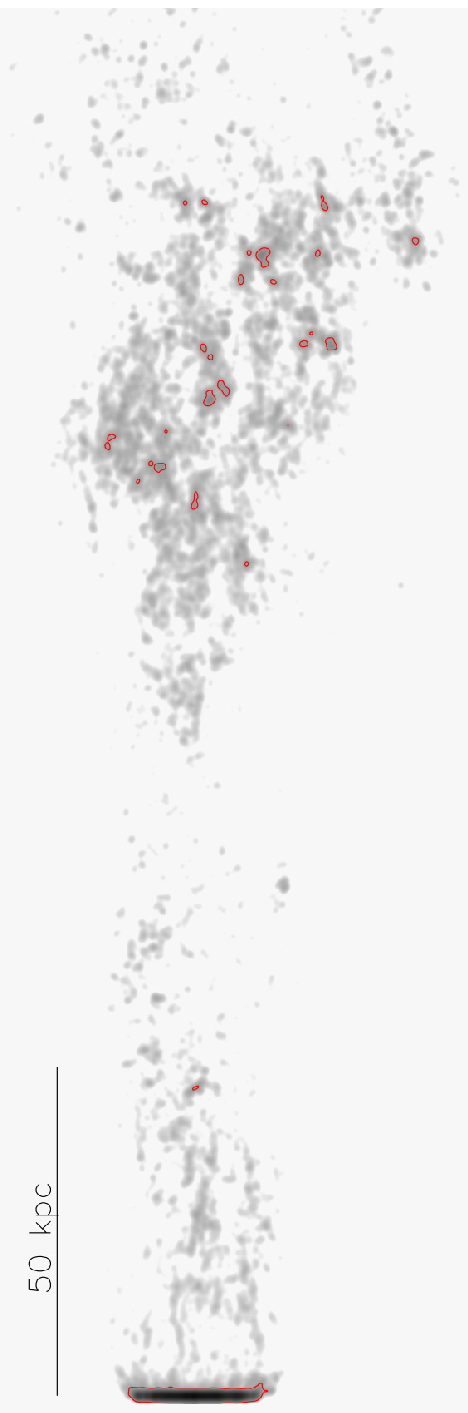}
\includegraphics[scale=0.8, trim= 17mm 0mm 13mm 0mm, clip]{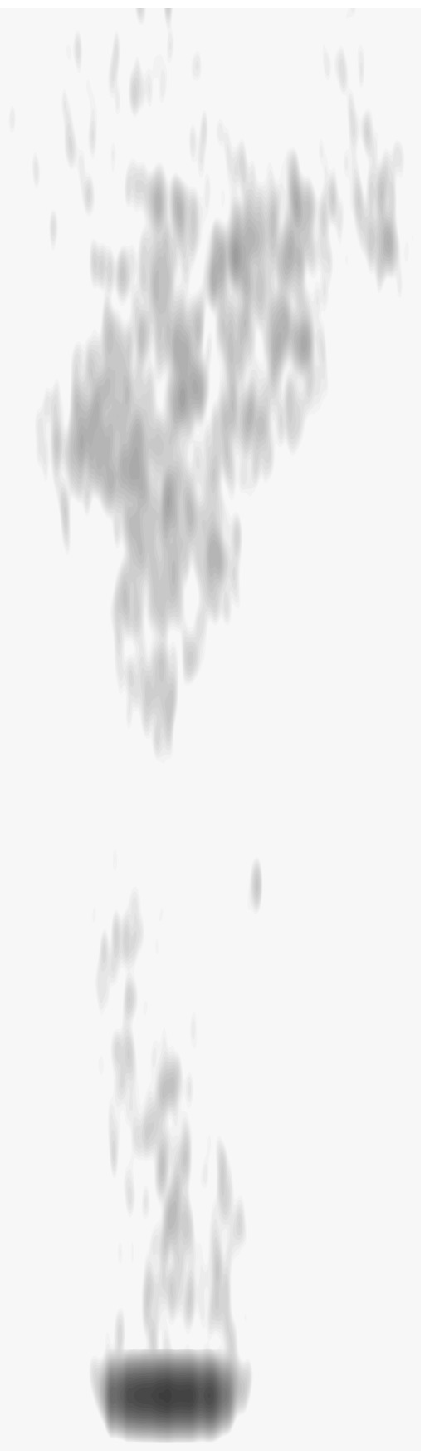}
\includegraphics[scale=0.8, trim= 17mm 0mm 13mm 0mm, clip]{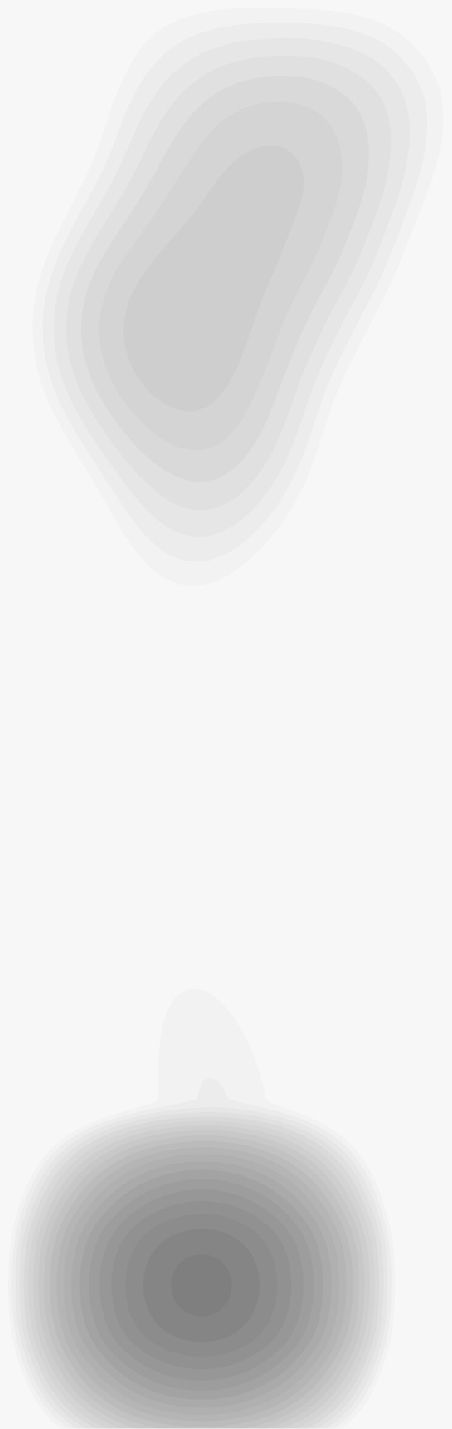}
\includegraphics[scale=0.7]{f6d}\\
\end{center}
\caption{The column density of \ion{H}{I} gas observed at a resolution of 1.2 kpc (to compare with Chung et al. 2007), 1.4 kpc $\times$ 7.4 kpc (to compare with Oosterloo \& van Gorkom 2005), and 43 kpc (to compare with Vollmer \& Hutchmeier 2007).  These projections are 500 Myr after the wind has hit the galaxy for the standard case with cooling to 8000 K.  The greyscale shows the log of the column density in units of cm$^{-2}$.  We add a red contour to the left panel, denoting a column density of 3 $\times$ 10$^{20}$ cm$^{-2}$, a lower limit for the column density of gas associated with star formation (Schaye 2004). }
\label{fig:hicomp}
\end{figure*}

Figure~\ref{fig:hi} is the \ion{H}{I} map from the $\tmin = 8000$ K run.  The difference between the $\tmin = 8000$ K and $\tmin = 300$ K runs (not shown) is minor, similar to the density distribution (see Figure \ref{fig:gas}).  The $\tmin = 300$ K run generally has a few more clouds, with slightly higher \ion{H}{I} densities, where the total gas surface density is highest.  The gap seen in the $\tmin = 8000$ K run is at the same height above the disk, but smaller in the $\tmin = 300$ K run.  At late times (600 Myr after the wind has hit the galaxy), the width of the \ion{H}{I} tails is about the same in the two runs, less than 40 kpc wide, and there is a paucity of high surface density clouds in both runs.  

The observations in Figure~\ref{fig:hi} are at unrealistically high resolution, even for \ion{H}{I} interferometric maps.  We can also ``observe" our simulation by smoothing to lower resolutions, like those in typical observations; this is done in Figure \ref{fig:hicomp}.  The first panel has a resolution of 1.2 kpc, similar to the observations of Chung et al. 2007.  The middle panel has a resolution of 1.4 kpc $\times$ 7.4 kpc (to compare with Oosterloo \& van Gorkom 2005), and the right panel has a resolution of 43 kpc (to compare with Vollmer \& Huchtmeier 2007).  As our smoothing size increases we find fewer clouds because some of our high density clouds are small enough that they are smoothed below our threshold column density of 10$^{19}$ cm$^{-2}$.  These three projections are all at 500 Myr after the wind has hit the galaxy, and are all from the $\tmin = 8000$ K run.  As in Figure \ref{fig:hi}, the lowest contour is at the minimum column density of current observations, 10$^{19}$ cm$^{-2}$.


We begin by comparing our left panel in Figure~\ref{fig:hicomp} to the results of Chung et al. (2007), finding that our tails are generally longer.  However, our tail is from a galaxy that has been stripped for 500 Myr, while the authors argue that the tails they observe are in the early stages of stripping.  Indeed, about 50 Myr after the wind hits (see the left panel of Figure~\ref{fig:hi}) our tail is well within the range observed in these outer Virgo cluster galaxies.  Also, by examining their Figure 2, we see that the estimated ram pressures of all but one of the observed galaxies are less than our simulated ram pressure, so the tails will lengthen more slowly.  Finally, our measured tail length is the actual length, while the tails observed by Chung et al. (2007) are likely to be projected at an angle to the line of sight.  

Next, we compare our results to the observations of Oosterloo \& van Gorkom (2005), who observe an \ion{H}{I} tail 110 $\times$ 25 kpc in size, and estimate the tail to be a few hundred Myr old.  See the central panel in Figure \ref{fig:hicomp}.  In either of our cooling cases, we can observe tails longer than this one 500 Myr after the wind has hit the galaxy.  Interestingly, in the central panel, there is a gap with no emission, just as is seen in the tail of NGC 4388.  The range of surface densities in the tail in this panel are similar to those found by Oosterloo \& van Gorkom (2005), although our maximum \ion{H}{I} surface density is about a factor of two less than they observe.  In the $\tmin = 300$ K case at this resolution, there is a clump with a surface density above 3 $\times$ 10$^{20}$ cm$^{-2}$, the star-forming lower limit, 200 kpc from the disk 500 Myr after the wind has hit.  

This tail of \ion{H}{I} clouds would not have been found by Vollmer \& Hutchmeier (2007), as we show in the right panel of Figure \ref{fig:hicomp}.  At this very low resolution of about 43 kpc, only the more distant gas in the tail is observable, and is more distant from the galaxy than the $\sim$60 kpc search radius adopted in that paper.  

We find that the ability of the gas to radiatively cool is vital to the existence \ion{H}{I} tails.  As we discuss in Section \ref{sec:rhot}, as long as radiative cooling in the stripped gas is more efficient than adiabatic and turbulent heating, \ion{H}{I} clouds will exist in our simulated tails.  If heat conduction is also not efficient enough to counteract radiative cooling, long \ion{H}{I} tails consisting of dense clouds should be fairly common. 

In both of the cooling runs, the tail has clouds that are above the column density of disk gas that is associated with star formation, 3 $\times$ 10$^{20}$ cm$^{-2}$ (Schaye 2004).  If these clouds could form stars, ram pressure stripped gas  could result in the trails of ``fireballs" observed by Yoshida et al. (2008) in RB 199.  

\subsection{H$\alpha$}\label{sec:halpha}

Next, we turn to H$\alpha$ emission.  The minimum observable surface brightness that we use for H$\alpha$ emission is 2 $\times$ 10$^{-18}$ erg s$^{-1}$ cm$^{-2}$ arcsec$^{-2}$ (see Sun et al. (2007) and references therein).  Similar to our calculation of hydrogen neutral fraction, we use Cloudy to determine the H$\alpha$ emissivity given the gas temperature and density.  

\begin{figure}
\includegraphics[scale=0.65,trim= 13mm 11mm 13mm 110mm, clip]{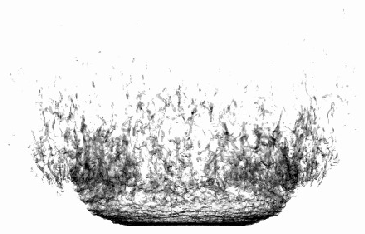}
\includegraphics[scale=0.65,trim= 13mm 11mm 13mm 110mm, clip]{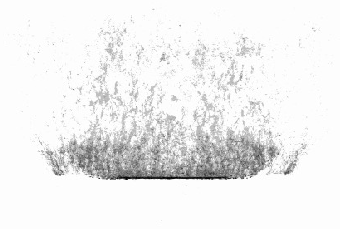}\\
\includegraphics[scale=0.65,trim= 13mm 11mm 13mm 45mm, clip]{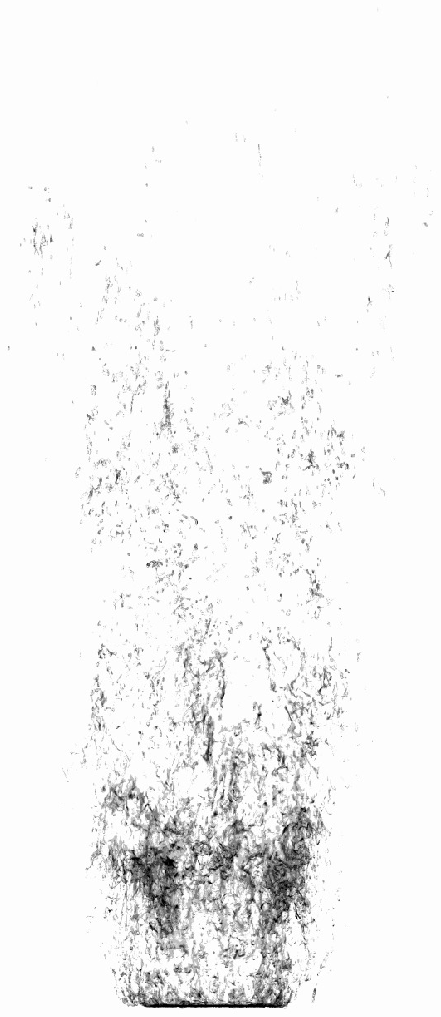}
\includegraphics[scale=0.65,trim= 13mm 11mm 13mm 45mm, clip]{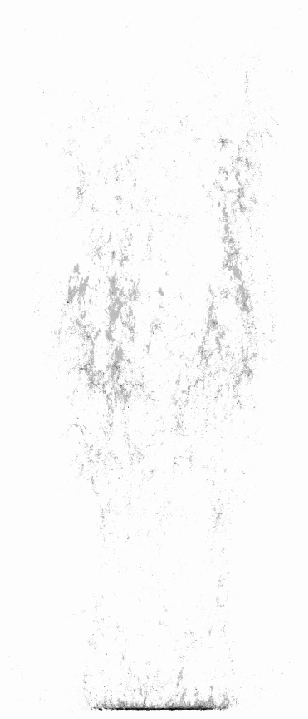}\\
\includegraphics[scale=0.65,trim= 13mm 11mm 13mm 0mm, clip]{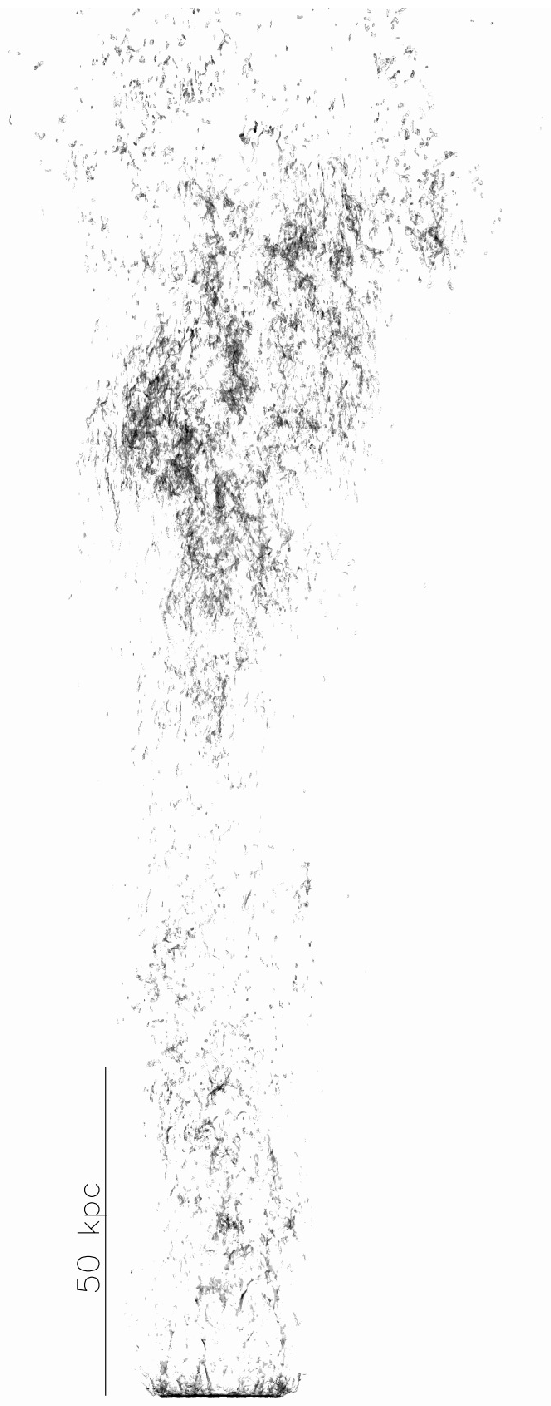}
\includegraphics[scale=0.65,trim= 13mm 11mm 13mm 0mm, clip]{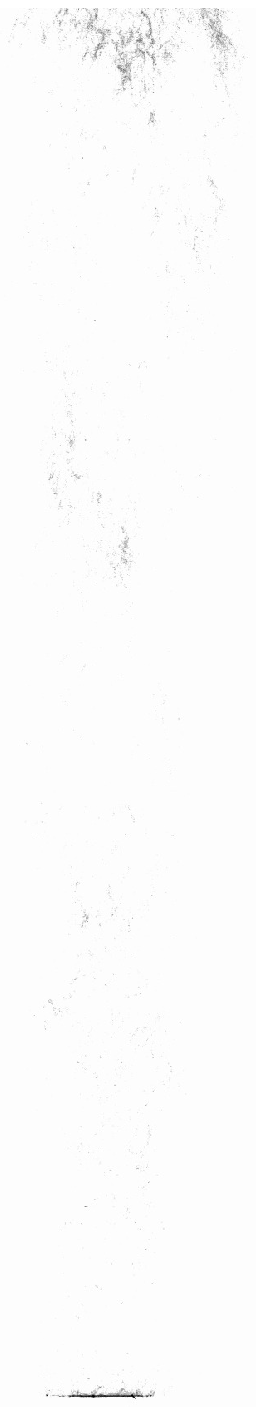}
\includegraphics[scale=0.4,trim=0mm 0mm 0mm 7mm,clip]{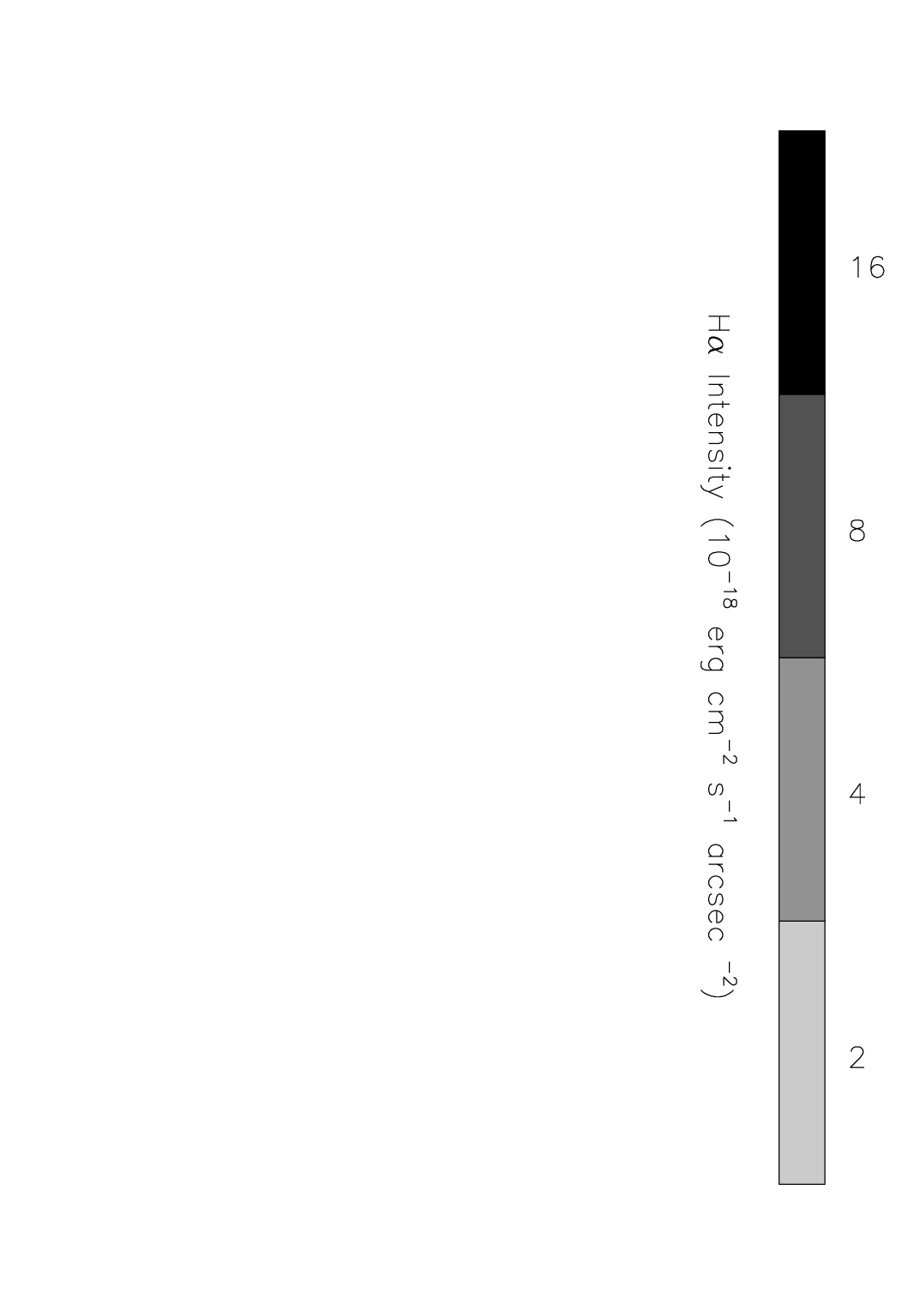}\\
\caption{The H$\alpha$ surface brightness contours that  could be observed using current observational depths (Sun et al. 2007).  Again, these are for 100 Myr, 250 Myr, and 500 Myr after the wind has hit the galaxy.  The left column is the simulation with radiative cooling to 8,000 K and the right side has cooling to 300 K.  Note that the H$\alpha$ emission is filamentary in structure and follows the \ion{H}{I} column density.  See section \ref{sec:halpha} for discussion.}
\label{fig:halpha}
\end{figure}

We find, as shown in Figure \ref{fig:halpha}, highly structured, long tails of H$\alpha$ emission.  Note that we do not include UV radiation from star formation or AGN (except for the metagalactic background, as described in section \ref{sec:projection}).   The H$\alpha$ emission roughly follows the \ion{H}{I} column density, which is likely where the gas is densest and where competition between compressive and turbulent heating and radiative cooling is taking place, as discussed in Section \ref{sec:rhot}.  The $\tmin=8000$ K case shows significantly more H$\alpha$ emission than the $\tmin = 300$ K run, which is due to two reasons: first, the gas is kept close to the temperature where collisional excitation can result in emission.  The other difference is that in the $\tmin=300$ K simulation, much of the gas is at higher densities, where it is optically thick to the metagalactic ionizing background and so produces little H$\alpha$ emission (and it is not hot enough to emit due to collisional excitation).

Turning to observations, it is important to note that many of the observations of H$\alpha$ emission are closely linked to star formation or AGN activity in a galaxy (which we do not include in the simulations).  However, in the case of NGC 4438 (Kenney et al. 2008), the emission is not close to either of these ionizing sources.  The brightness of the filaments matches well with the H$\alpha$ emission in our simulation with radiative cooling to 8000 K, with the lowest observed surface brightness emission at the level of our third contour.  The overall filamentary structure and large gaps between emission regions in our projected H$\alpha$ emission also matches the observations of Kenney et al. (2008).

Gavazzi et al. (2001) see emission slightly above the brightness of our third contour, although it is difficult to discern the structure of the observed tails.  They consider it likely that the gas was ionized within the galaxy and has remained ionized in the tail, although they do not rule out heating within the tail.  We believe that turbulent and adiabatic heating of stripped cool gas causes much of the emission we observe in our projections, because of the close alignment of the \ion{H}{I} and H$\alpha$ emitting gas added to our earlier discussion of differential acceleration (Section \ref{sec:vel}).  

As discussed in section~\ref{sec:projection}, we computed our \ion{H}{I} and H$\alpha$ maps using Cloudy, assuming a typical cloud thickness of 100 pc, which is appropriate for our resolution.   To explore our sensitivity to this assumption, we also generated maps based on cloud size of 10 pc (the `thin' case), and 1000 pc (the `thick' case).  

The thick case is very similar to our standard 100 pc cloud size, with slightly more \ion{H}{I} in both of our cooling runs.  In the $\tmin = 8000$ K run, the H$\alpha$ emission is nearly identical, possibly because collisional excitation dominates the emission.  In the $\tmin = 300$ K run, the thicker clouds emit a little less H$\alpha$ at our lowest observable brightness -- although the second contour level is nearly identical.  

The thin cloud approximation does result in lower \ion{H}{I} column densities, but in both of the cooling cases there are still dense clouds at star forming column densities 500 Myr after the wind hits the disk.  The H$\alpha$ emission is considerably brighter, by about a factor of 8 in the $\tmin = 8000$ K run.  However, in the $\tmin = 300$ K run, the emission is much brighter, frequently by a factor of about 16. 

\begin{figure*}
\begin{center}
\includegraphics[scale=0.65,trim=5mm 0mm 5mm 0mm,clip]{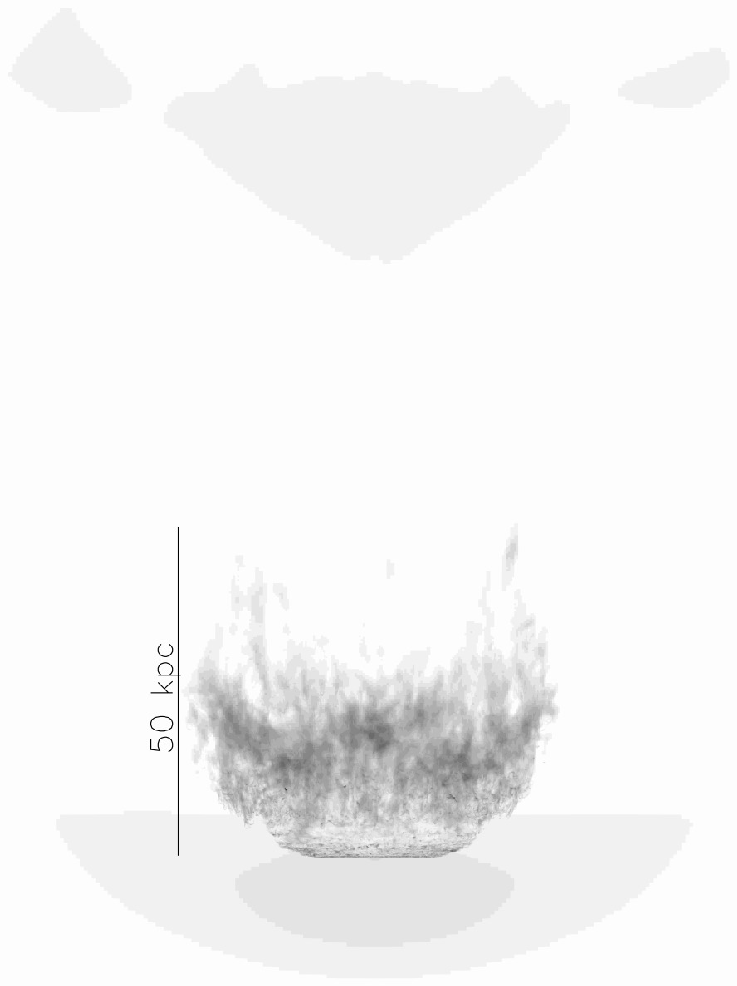}
\includegraphics[scale=0.65,trim=5mm 0mm 5mm 0mm,clip]{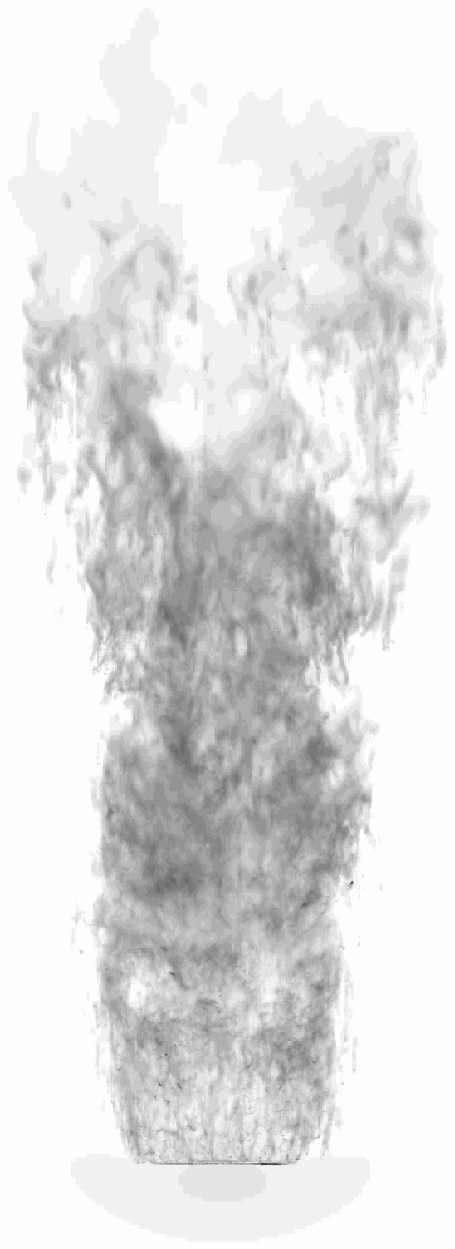}
\includegraphics[scale=0.65,trim=5mm 0mm 5mm 0mm,clip]{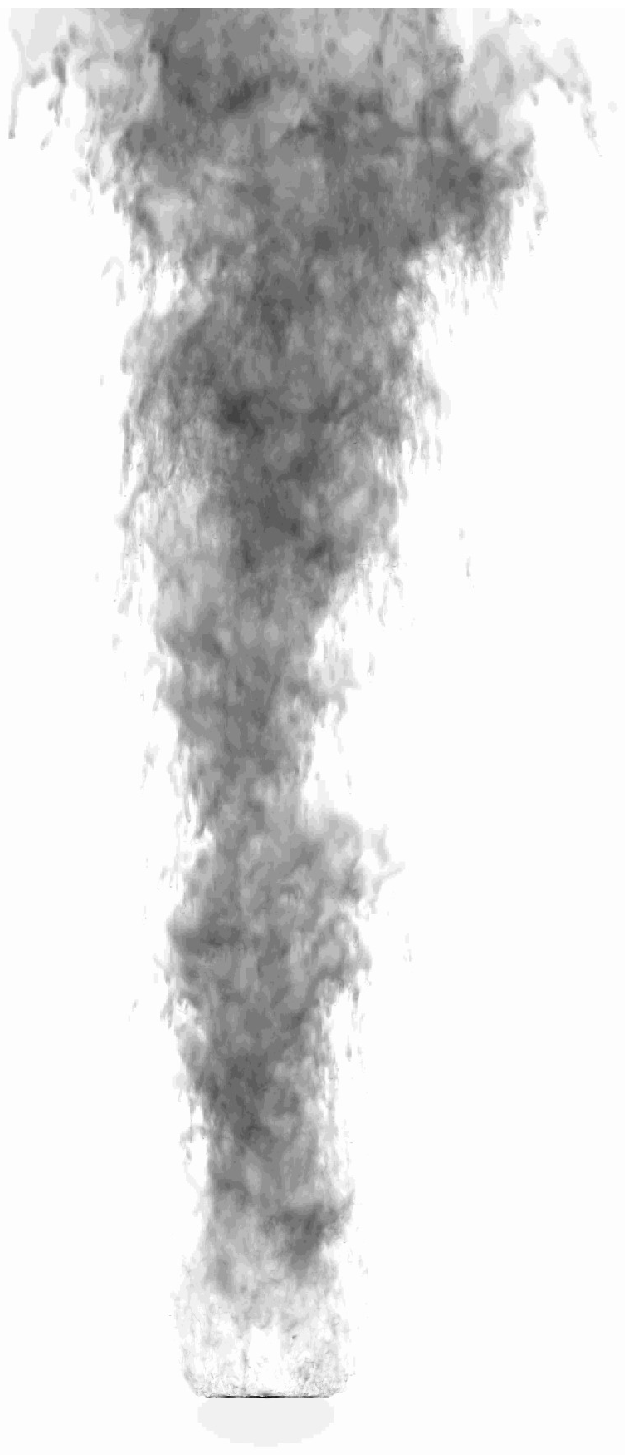}
\includegraphics[scale=0.7]{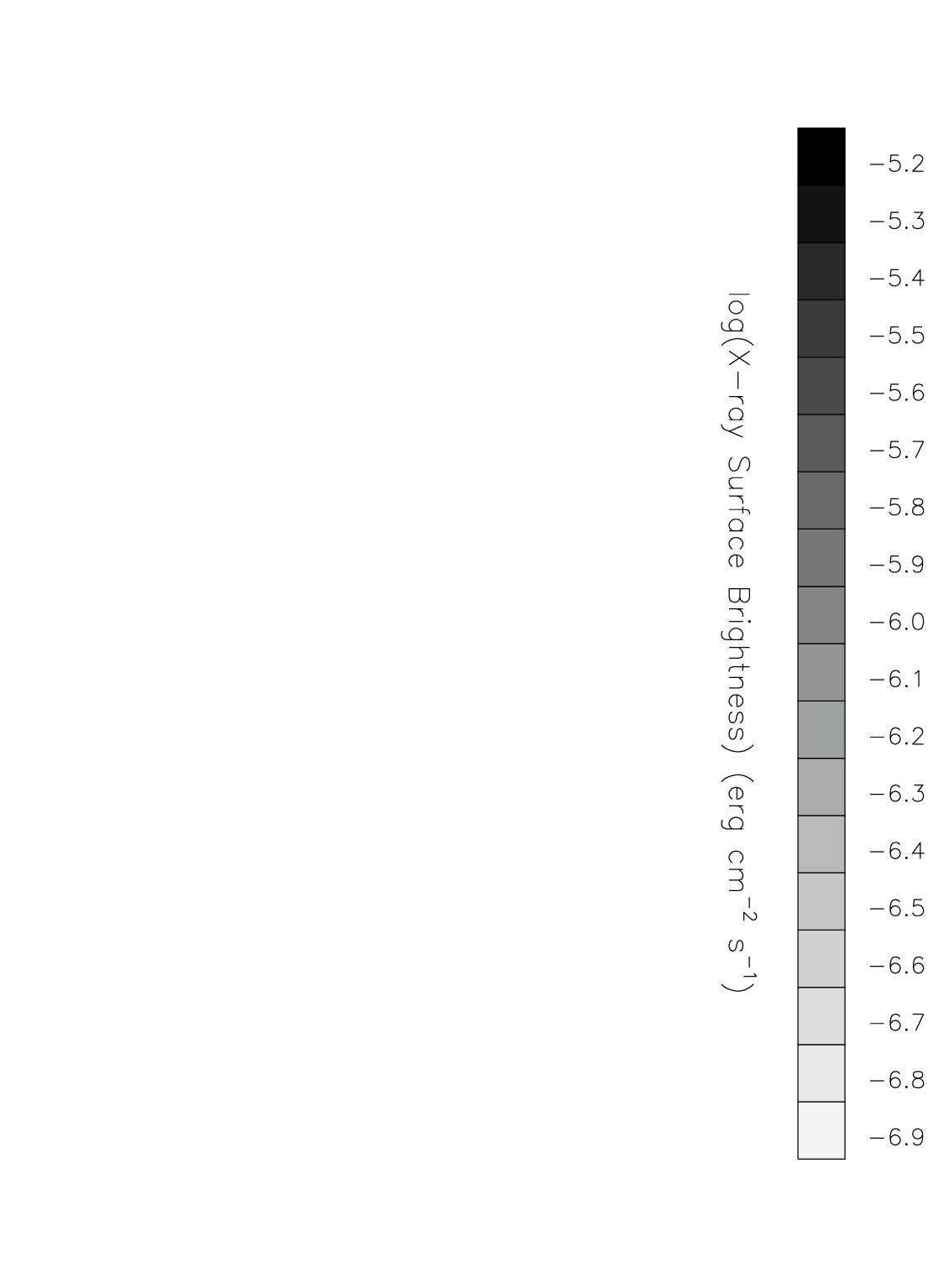}
\end{center}
\caption{The predicted X-ray surface brightness in the 0.5-2 keV band for the $\tmin = 8000$ K run at 100 Myr, 250 Myr and 500 Myr after the wind has hit.  Note that none of the gas in the tail is bright enough to be observed at current observational limits. }
\label{fig:xray}
\end{figure*}

Another concern is that, by using Cloudy in this way to generate our emissivities, we are assuming ionization equilibrium.  Given the short cooling time, it is possible to be out of equilibrium which could systematically bias the emission.  To check how robust this approximation is, we used the simulation which followed the non-equilibrium ionization fractions of H and He.  We computed the emissivity two ways: first, with our standard Cloudy procedure which uses only the total density and temperature of the gas, and second, based on the actual ionization fractions predicted in the simulations based on a set of analytic fits to the H$\alpha$ emissivity due to recombination (Spitzer 1978), and the H$\alpha$ emission due to collisional excitation (Osterbrock 1989).   We found that both methods produced very similar H$\alpha$ maps (although note that this run did not include cooling below $T = 10^4$ K).

\subsection{X-ray}\label{sec:xray}

Finally, we turn to X-ray emission.  As noted in the introduction, only a few X-ray tails are known.  We show predicted X-ray images in the 0.5-2 keV band in Figure~\ref{fig:xray}.  None of the contours are bright enough to be observed assuming an observational limit of 5 $\times$ 10$^{-6}$ erg s$^{-1}$ cm$^{-2}$ (based on the luminosity and surface area of the tail in Sun et al. 2006).  A bow shock is visible early in the simulations, but most of the emission comes from gas in the wake.  Lowering the radiative cooling floor to 300 K does not effect the X-ray surface brightness substantially.

There are a number of possible difference between our simulation and the X-ray observations of tails.  First, since we do not include energetic feedback from supernovae,  the gas in the tail that originated from the galaxy could not have been heated by supernovae explosions, as suggested by Sun \& Vikhlinin (2005).  

However, in the case of ESO 137-001, Sun et al. (2006) conclude that the X-ray emission they observe is from mixing of the cool galactic gas with the hot ICM.  This type of heating leading to X-ray emission could in fact be happening in our simulation, but just does not lead to observable emission.  The ICM in our simulation is about 60\% of the temperature and a third of the density of the ICM in A3627.  Our galaxy velocity is 1.4 times larger than the estimated velocity of ESO 137-001, so our ram pressure is ~63\% as large as that experienced by the observed galaxy.  However, at a distance of 200 kpc from the cluster center, a galaxy velocity of 2000 km/s is much more likely (Tonnesen \& Bryan 2008), which results in our ram pressure being only 16\% of that experienced by ESO 137-001.  Based on their fits, the authors predict an X-ray tail temperature of ~9$\times$10$^6$ K and a density of ~10$^{-26}$ g cm$^{-3}$.  We have no tail gas from our galaxy in any of our simulations at that temperature and density.  However, if we shift our contour plots in Figure \ref{fig:rhot} by the ratio of the ICM pressure of A3627 and our simulated cluster, a small amount of gas is at the required density and temperature.  This suggests that the ICM conditions may be important for producing X-ray bright tails.


\section{Discussion}\label{sec-discussion}

As we have demonstrated in the previous section, our stripping simulations with cooling result in very different wakes than in runs without cooling.  This impacts both the morphology of the tail, the density and temperature structure of the gas, and the observational diagnostics.  In this section, we begin by discussing one point which we have noted throughout: the differences between the $\tmin = 8000$ K and $\tmin = 300$ K simulations.  Then we investigate the effect of resolution and finally turn to a discussion of the physical mechanisms which we do not include in this simulation.

\subsection{Radiative Cooling Floor}\label{sec-cool}

As noted earlier, we used two different values for our radiative cooling floor (8000 K and 300 K) in order to explore in a simple way the potential impact of processes which we do not include in the simulation.  We found that most of our predictions are insensitive to this parameter, which gives us confidence in our results.  However, there were some differences between the runs.  This included the detailed density-temperature structure of gas in the tail (at the low-T end).  The biggest difference was the amount of H$\alpha$ emission, and the survival of \ion{H}{I} clouds, with $\tmin = 300$ K predicting longer lived clouds and less H$\alpha$ emission.

It is unclear which $\tmin$ value is preferable.  Because of the lower Jeans length in the lower temperature floor runs, we resolve the resulting fragmentation less well in the run with a minimum temperature of 300 K.  Although the $\tmin = 300$ K run is desirable because it allows for the existence of the low temperatures that are found in the ISM, it is not clear if this run is physically more realistic as we do not include some effects such as small-scale turbulence, cosmic rays and magnetic fields, all of which may provide a source of effective pressure in low temperature regions.  

\subsection{Resolution}\label{sec:res}

We now discuss the impact of resolution on our results.  To do this, we can compare our standard simulations with runs with lower mass resolution, resulting in less refinement in the tail.  With either refinement criteria, the gas in the disk is mostly refined to a resolution of 38 pc when the winds hits the disk, but in the case of more mass refinement, the disk fragmentation is better resolved, resulting in a larger number of lower mass clumps.  This affects the evolution of the disk.  In our radiatively cooled cases, the gas fragments from the edge inward.  With lower mass resolution, the resulting clouds are more massive and can perturb the inner disk, causing fragmentation and the formation of over- and under-dense regions.  In contrast, in the highly resolved cases that we discuss in this paper, the disk has fewer holes in the central disk, and therefore stripping does not occur in the central 7 kpc.  Also, stripping occurs more slowly, although a slightly larger amount of gas is stripped from the galaxy (this is very similar to our discussion regarding the case with cooling to 300 K in TB09).    

With the lower mass resolution run, most of the gas in the tail is refined to either 304 pc or 152 pc, and none of the tail is refined to 38 pc.  This is in contrast to the more refined tail, in which most of the gas is refined to either 152 pc or 76 pc, and there are a number of small areas refined to 38 pc.  We go through our results discussed in this paper, highlighting the similarities and differences of different resolution runs.

\begin{figure}
\includegraphics[scale=0.55,trim=5mm 0mm 5mm 0mm,clip]{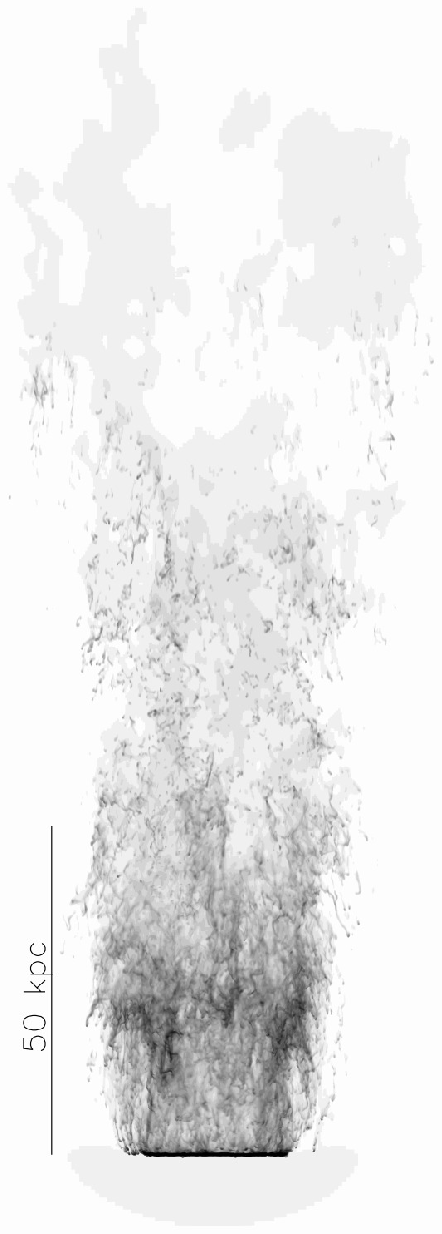}
\includegraphics[scale=0.55,trim=5mm 0mm 5mm 0mm,clip]{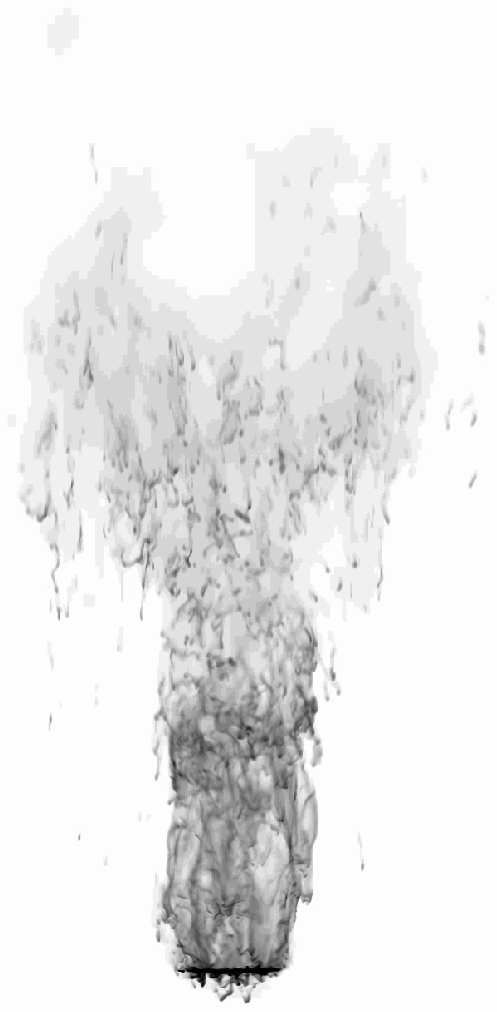}
\caption{Contour plots of total gas surface density, comparing two $\tmin = 8000$ K cases with different mass resolution 250 Myr after the wind has hit.  The left figure is our standard high resolution run, and the right is the lower mass resolution run.  Because there is less gas clumping in the disk, especially in the outer regions, the lower resolution run has a smaller remaining disk.   There is more fallback in the lower resolution run, resulting in some of the gas punching back through the disk.  Although the dense clouds in either case are close to the same distance above the disk, the less dense gas has clearly been accelerated more quickly from the more highly refined disk. }
\label{fig:gasres}
\end{figure}

The morphology of the stripped gas differs early in the stripping process, but by 500 Myr after the wind has hit the galaxy, the stripped gas looks very similar.  In Figure \ref{fig:gasres} we show the most different projection of either of our comparison sets of runs ($\tmin = 8000$ K and $\tmin = 300$ K), using the $\tmin = 8000$ K cases at 250 Myr after the wind has hit the galaxy.  The lower mass resolution case has fewer dense clouds at the edges of the disk, so most of that gas is quickly stripped.  However, the higher resolution run has a slightly longer tail because the very low density gas in that run has been accelerated more quickly to larger distances.  The galaxies with more refinement have more and smaller clouds in the tail, as expected.

The gas density and temperature distributions are very similar.  Although the gas distribution remains similar, at late times the runs with lower resolution in the tail have less high density gas and a slightly lower maximum density.  This is because clouds do not condense to as small sizes in the lower resolution tail.  

The tail velocity profiles are also very similar, with similar widths and the same maximum velocities reached at the same distance from the galactic disk.  All radiatively cooled runs have a consistent velocity width over time.  There is also some fallback onto the disk in the lower resolution run, because larger gas clouds in the disk cause fragmentation in the central regions and therefore also result in gas stripping in those regions.  This centrally stripped gas can then be protected by disk gas that is rotating or that moves to fill the center of the disk (discussed in detail in TB09), and will be shielded from the stripping wind.  This protection allows even more dense gas to have slightly negative velocities in the lower resolution cases.  Thus, the gas punching through the bottom of the disk in our lower resolution runs and in the Kapferer et al. (2009) simulations may be resolution dependent.

\begin{figure}
\includegraphics[scale=1.1, trim= 23mm 12mm 23mm 0mm, clip]{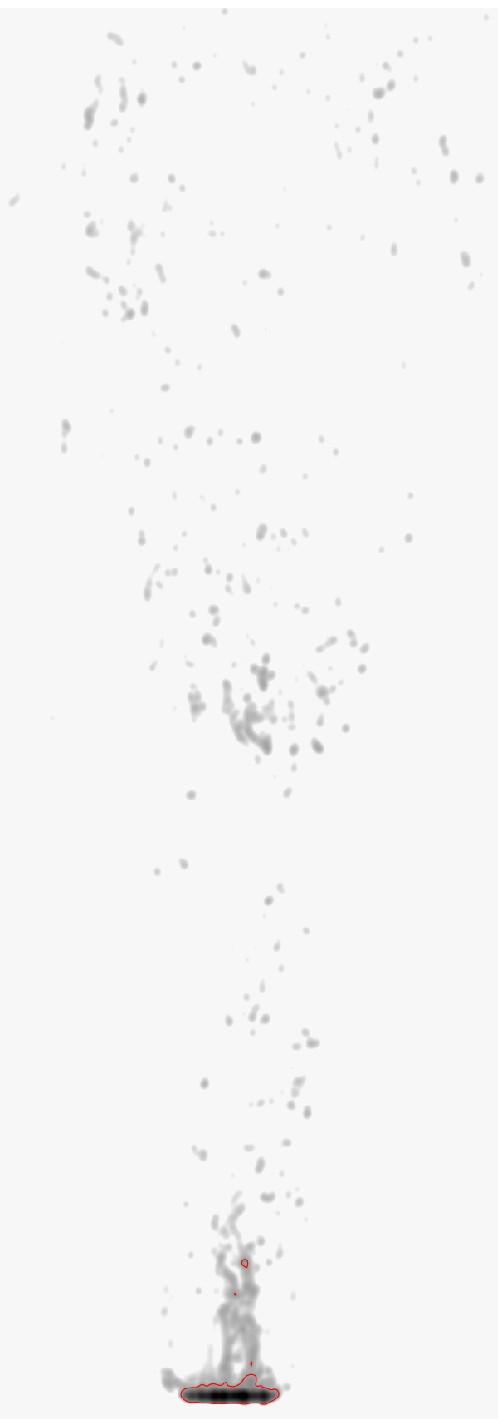}
\includegraphics[scale=1.1, trim= 23mm 12mm 23mm 0mm, clip]{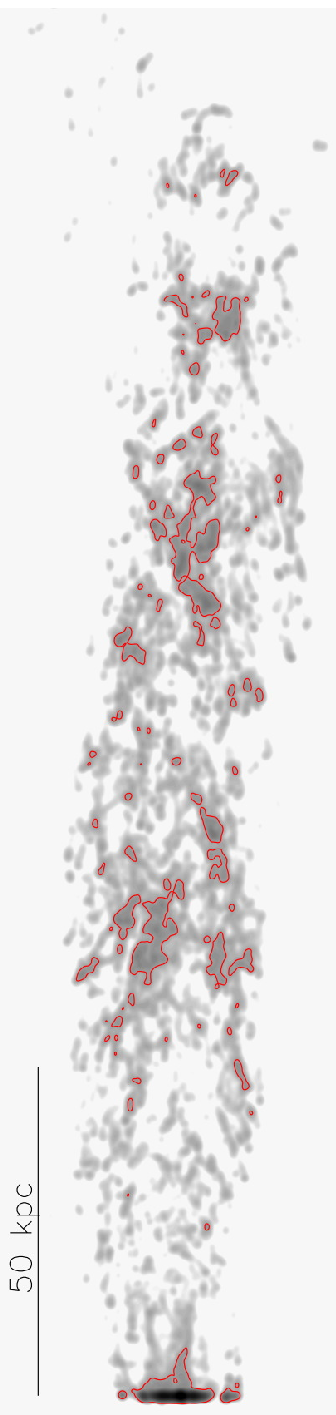}
\caption{Contour plots of \ion{H}{I} gas column density at 1.2 kpc resolution, comparing the two lower mass resolution runs with radiative cooling.  These are both projected 500 Myr after the wind has hit the galaxy, and can be compared to the left panel of Figure \ref{fig:hicomp}.  $\tmin = 8000$ K is on the left, and $\tmin = 300$ K is on the right.  Notice that there is more dense gas closer to the disk.  As we discuss in Section \ref{sec:res}, centrally stripped gas is able to sit in the protected lee of the disk, and dense gas does not survive as far from the disk because it does not collapse into as small dense clouds.  There is more dense gas in the $\tmin = 300$ K case because the lower cooling floor results in smaller clouds.}
\label{fig:hires}
\end{figure}

The \ion{H}{I} maps are somewhat different, again because of the increased fragmentation in our highly resolved tails.  Compare the two projections in Figure \ref{fig:hires} to the first panel in Figure \ref{fig:hicomp}.  We find that the $\tmin = 300$ K run produces more \ion{H}{I} in our maps than the $\tmin = 8000$ K run, a more marked difference than in our standard high mass resolution runs.  This is because when we use the more stringent mass criterium to resolve the tail, only the lower cooling floor creates high density clouds that are above our refinement threshold and therefore more highly resolved.  The \ion{H}{I} tails are generally shorter, and the $\tmin = 8000$ K run has much less gas at large distances in the tail.  Also, both cooling floors have tails that are more connected to the disk in the lower resolution runs.  This occurs for two reasons: first, the tail is shorter because the gas is less dense and is more easily heated by the turbulence in the tail.  Second, gas can be stripped from the central region of the disk and then protected by rotating or inflowing disk gas.  Therefore, gas that is immediately behind the disk may be protected from the wind and can survive there for longer.

H$\alpha$ emission is largely unaffected, because the same type of turbulence exists in the different resolution cases despite the difference in size of the dense clouds.  The interaction of the slower moving dense clouds with the faster low density wind still causes small scale turbulence that can heat the nearby gas.  The difference in resolution does not change the brightness of the emission because the temperatures are the same, resulting in similar amounts of collisional excitation.  If the difference in clouds sizes were enough to greatly affect the optical depth of the clouds, H$\alpha$ emission would decrease.  The main difference is the distribution of emission, which follows the \ion{H}{I} clouds in the tail in any run.  Therefore, as with the \ion{H}{I} gas, the H$\alpha$ emission is closer to the disk.  As we discussed above, the cooling floor has a much larger effect.

The X-ray emission is affected very little.  As we discussed in Section \ref{sec:xray}, the surrounding ICM pressure likely needs to increase for X-ray emission to be observable. 

In summary, the different results caused by using different resolutions can be attributed to two main causes.  First, and most importantly, less resolution in the tail results in less fragmentation to smaller dense clouds (although there is more fragmentation in the $\tmin = 300$ K run than in the $\tmin = 8000$ K run).  Second, gas stripped from the galactic center can be protected by shadows from the wind by the remaining disk gas, and fall back towards the disk. 

\subsection{Heat Conduction}

Heat conduction could be important for the survival of cool clouds in the ICM and for H$\alpha$ emission.  If heat conduction is an efficient way to transport heat from the ICM to cold, stripped gas, then the survival time of \ion{H}{I} clouds would be less than predicted in this paper, and the length of the tails would be shorter.  The fact that the length of our tails matches observations of either shorter tails by Chung et al. (2007) for our $\tmin = 8000$ K case, or long tails like that observed by Oosterloo \& van Gorkom (2005) using the $\tmin = 300$ K simulation, argues against a strong role for heat conduction, at least for the current conditions.

We can perform an estimate of the evaporation time for a typical cloud if heat conduction is not suppressed.  We follow Cowie \& McKee (1977), as in Vollmer et al. (2001).  In our clouds, we find that the mean free path for ions is comparable to or greater than the temperature scale length, so we need to use the saturated heat flux equations.  Solving for the evaporation time, we find that using a cloud radius of 100 pc, and a number density of about 0.1 cm$^{-3}$, t$_{evap}$ $\sim$ 6 $\times$ 10$^5$ years.  However, this time can be lengthened if there are magnetic fields perpendicular to the temperature gradient (Cowie \& McKee 1977; Cox 1979).  Vollmer et al (2001) argue that magnetic fields could increase the evaporation time by nearly an order of magnitude, although the actual value is highly uncertain.  Although McKee \& Cowie (1977) predict that a cloud in our modeled ICM would be evaporated before being radiatively cooled, the cooling time for a cloud of density  0.1 cm$^{-3}$ and temperature 10$^5$ K is 10$^6$ years.  A higher density shortens the cooling time and lengthens the evaporation time, so radiative cooling may also slow down the effects of heat conduction.

\subsection{Magnetic Fields}

Our simulation does not include magnetic fields or cosmic rays.  In general in the Milky Way, the magnetic field is measured to be about 2-3 $\mu$G (Men et al 2008).  However, there are measurements of the magnetic field in molecular clouds (Crutcher 1991 and references therein) of up to $10^3$ $\mu$G.  If some of the denser clouds we see in the tails in our simulations had strong magnetic fields, these fields could protect the clouds from ablation while they move through the wind, in addition to lowering the efficiency of heat conduction.  We do not know to what extent magnetic fields could counteract turbulent heating and ablation of stripped clouds because we do not know the amount of structural support magnetic fields would give stripped clouds. 

The first long stripped tails were observed in radio continuum, with stretched magnetic field lines downstream of the galaxy (Gavazzi et al. 1995).  Including magnetic fields in our simulation would likely also result in radio continuum emission (assuming a cosmic ray population was also stripped), although because our ram pressure is less than the ram pressure estimated to be affecting the three galaxies in A 1367, the magnetic fields would be stretched less.

\subsection{Star Formation}\label{sec-sf}

A molecular cloud may survive between a few and a few tens of Myr (Blitz \& Shu 1980; Larson 2003; Hartmann 2003).  Because we do not include star formation, our dense clouds of gas do not evolve into less dense gas that could be stripped.  If our dense regions of gas could turn into low density regions through star formation ($\rho$ $\le$ 10$^{-23}$ g cm$^{-3}$), our galaxies that include radiative cooling could be entirely stripped of gas over the lifetime of the dense clouds.  This could result in a tail without a gap between the majority of the stripped gas and the galactic disk.  

We find that overdensities in the stripped tails can survive to large distances, however, our projections also show that they are being ablated by the ICM wind.  Our next step is to add star formation, which could affect stripping in a number of ways.  Ram pressure could induce star formation in the disk, using up disk gas and resulting in lower mass tails and fewer dense clouds in the tail.  If there are dense clouds in the tail, we will discover whether stars can form in stripped gas, or if Kelvin-Helmholtz stripping will destroy clouds before they can collapse.

\section{Conclusions}

We have run detailed galaxy simulations including radiative cooling to understand the morphology of and emission from gas tails stripped by ram pressure in a cluster environment.  We compare three cases: a run without cooling, one with cooling to a minimum temperature floor of $\tmin = 8000$ K to approximately account for non-thermal pressure support, and one to $\tmin = 300$ K.  Our main conclusions are as follows:

1. Including radiative cooling creates a clumpy ISM, which results in faster stripping of lower density gas, and the production of longer tails.  A clumpy disk also has more holes for the ICM wind to pass through, which results in significantly less flaring in the stripped material, and narrower tails in the cooling runs, in better agreement with observed tails (e.g. Chung et al. 2007).

2. We find that the clumpy ISM also produces a much wider range of densities and temperatures in the wake, resulting in a very different morphology for the tail.   Although the stripped gas is mostly at the ICM pressure, it shows a wider distribution in pressure, as well as a systematic deviation from the constant pressure line for temperatures below $T = 10^5$ K, due to radiative cooling operating more quickly than compressive heating.

3.  The stripped gas in the cases with radiative cooling does not accelerate to the ICM wind speed.  This is likely because the dense clumps are harder to accelerate than a smooth distribution of gas.  Also, the gas is differentially stripped and differentially accelerated through the whole length of the tail for the duration of the simulation.    

4.  We compute realistic \ion{H}{I} maps of our tails and compare to \ion{H}{I} observations of stripped tails.  With either a radiative cooling floor of 8000 K or one of 300 K we find \ion{H}{I} tails with long lifetimes (more than 500 Myr), and large lengths (well over 100 kpc).  When we consider our run with cooling to 300 K, we find that the tails are longer and clouds can survive at high density for a longer time.  Most tails that have been observed in Virgo have shorter lengths, like our tails at very early times, however, many of these tails seem to be in the early stages of stripping (as in Chung et al. 2007), and may have only experienced a short amount of time at large ram pressure (see the orbits in Vollmer et al. 2001).  The few longer tails that have been observed are well matched by our results at late times in the cooling runs.  

5.  We can also map the H$\alpha$ emission from our runs, and find filamentary structures much like those found in deep observations of stripped tails (e.g. Sun et al. 2007).  Both the $\tmin = 8000$ K and the $\tmin = 300$ K case emit in H$\alpha$, but the simulation with the higher cooling floor has more emission and a higher maximum surface brightness in the tail.  The filamentary structure in either radiative cooling run reproduces the observations of the H$\alpha$ tail of NGC 4438  (Kenney et al. 2008), while the $\tmin = 8000$ K run has some emission that is bright enough to be observed at the Kenney et al. (2008) limit.  This H$\alpha$ emitting gas is near the \ion{H}{I} gas, indicating that it is created in the tail rather than surviving as H$\alpha$ emitting gas stripped from the disk.

6.  None of the simulated tails have X-ray surface brightnesses high enough to match observations of X-ray tails (note that some observed \ion{H}{I} or H$\alpha$ tails also do not show X-ray emission).  This may either be because we do not preheat the galactic gas with supernovae explosions, or that our ICM is not hot and dense enough for mixing to cause bright X-ray emission.

It is clear from our results that adding radiative cooling greatly improves the realism of simulated tails.  This is seen by comparing both the morphology and velocity structure of the simulated tails to observations.  We find, as in observations, that there are only two types of bright emission from our tail:  in our case \ion{H}{I} and H$\alpha$ emission.  It remains to be discovered whether all three types of emission can be at observable levels in a single tail or if the environment necessary for X-ray emission would destroy or heat the cooler clouds.  However, we can use our results and a few analytic estimates to make an initial guess:  if our ICM had the same pressure as that of A3627 we would likely have some gas in our tail with the necessary density and temperature for detectable X-ray emission (see Section \ref{sec:xray}).  At the same time, it may be that heating from compression or turbulence would still not be enough to heat clouds faster than they would be radiatively cooled, so we might still see \ion{H}{I} in the tail.  Finally, H$\alpha$ emission should be present as long as there is some intermediate density gas (near $T \sim 10^4$ K), so it may not be greatly affected by the ICM pressure.  To complete our comparison to the case of ESO 137-001, we have a lower ram pressure than that affecting the observed galaxy, and Sun et al. (2007) find that the ram pressure is likely to have triggered star formation throughout the disk.  This may have heated the cool gas clouds surrounding the star forming regions so much less cool gas was left to be stripped.  As we discuss in Section \ref{sec-sf}, we intend to add a star formation prescription to a future simulation, which will allow for a better understanding of how star formation can affect the gas tail. 

The \ion{H}{I} column density projections indicate that higher resolution observations (similar to that of Chung et al. 2007) of cluster galaxies would find more massive tails.  The fact that the \ion{H}{I} tails can have large gaps between the higher column density regions means that there could be tails that would not have been found by Vollmer \& Hutchmeier (2007), and points to a need for large surveys at the highest possible resolution.

Our simulations predict that only slightly deeper observations of tails in H$\alpha$ should reap great rewards of much more structure.  It is interesting that although less dense gas moves more quickly than the dense gas throughout the tail, the emission from H$\alpha$ is nearly perfectly aligned with the \ion{H}{I} column density, indicating that gas in our simulations is heated to H$\alpha$ emitting temperatures in-situ, rather than being stripped while ionized and surviving for hundreds of Myr.  

We can consider what observational signatures of ram pressure stripping our simulations predict are likely.  We find dense \ion{H}{I} clouds can survive far from the disk and for a significant amount of time.  This means that in order to determine that any \ion{H}{I} structure is not surviving stripped gas, other information such as kinematic data indicating rotation, is necessary (as in Haynes et al. 2007).  In addition, H$\alpha$ emission can be observably bright even without in-situ star formation.  However, the high column densities of some of the \ion{H}{I} clouds leads us to believe that stripped gas can condense to form stars that could be formed far from the galaxy, indicating that even a stellar component in a stripped tail is not enough to prove that a gravitational interaction took place.  

The high resolution combined with the addition of radiative cooling in our simulation allows us to follow the interaction of dense cool clouds in the tail with the hot gas of the ICM.  Unlike some previous work, we find ablation and destruction of the clouds are important, as they move within the wind, limiting their survival to less than found by Kapferer et al. (2009).  It is important to reiterate that because it is difficult to make a physical case for either cooling floor, we present both 8000 K and 300 K cooling floors.  Although many of the main results about the tail attributes are very similar between the two cases, they do result in meaningfully different observable profiles in H$\alpha$.  Finally, although these simulations are arguably the most realistic stripping simulations performed to date, we have discussed in detail the limitations involved in our work in \S \ref{sec-discussion}.  

\acknowledgements

We acknowledge support from NSF grants AST-05-07161, AST-05-47823, and AST-06-06959, as well as computational resources from the National Center for Supercomputing Applications.  We thank Jacqueline van Gorkom, Jeff Kenney and other members of the Virgo group for useful discussions, as well as Elizabeth Tasker for invaluable help setting up the initial conditions.


\begin{thebibliography}

\bibitem[Agertz et al.(2007)]{2007MNRAS.380..963A} 
Agertz, O., Moore, B., Stadel, J., Potter, D., Miniati, F., Read, J., Mayer, L., Gawryszczak, A.,
Kravtsov, A., Nordlund, A., Pearce, F., Quilis, V.,
Rudd, D., Springel, V., Stone, J., Tasker, E.,
Teyssier, R., Wadsley, J., and Walder, R., 2007, \mnras, 380, 963

\bibitem[Arnaud(1996)]{1996ASPC..101...17A} Arnaud, K.~A.\ 1996, 
Astronomical Data Analysis Software and Systems V, 101, 17 

\bibitem[Blitz 
\& Shu(1980)]{1980ApJ...238..148B} Blitz, L., \& Shu, F.~H.\ 1980, \apj, 238, 148 

\bibitem[Bryan(1999)]{1999CoScE...1...46B} Bryan, G.~L.\ 1999, 
Comput.~Sci.~Eng., Vol.~1, No.~2, p.~46 - 53, 1, 46 

\bibitem[Burkert(1995)]{1995ApJ...447L..25B} Burkert, A.\ 1995, \apjl, 447, L25 

\bibitem[Byrd \& Valtonen(1990)]{1990ApJ...350...89B} Byrd, G., \& Valtonen, M.\ 1990, \apj, 350, 89 

\bibitem[Cayatte et al.(1990)]{1990AJ....100..604C} Cayatte, V., van Gorkom, J.~H., Balkowski, C., \& Kotanyi, C.\ 1990, \aj, 100, 604 

\bibitem[Chung et al.(2005)]{2005ASPC..331..275C} Chung, A., van Gorkom, 
J.~H., Kenney, J.~D.~P., \& Vollmer, B.\ 2005, Extra-Planar Gas, 331, 275 


\bibitem[Chung et al.(2007)]{2007ApJ...659L.115C} Chung, A., van Gorkom, 
J.~H., Kenney, J.~D.~P., \& Vollmer, B.\ 2007, \apjl, 659, L115 

\bibitem[Cowie 
\& McKee(1977)]{1977ApJ...211..135C} Cowie, L.~L., \& McKee, C.~F.\ 1977, \apj, 211, 135 


\bibitem[Cox(1979)]{1979ApJ...234..863C} Cox, D.~P.\ 1979, \apj, 234, 863 


\bibitem[Crowl et al.(2005)]{2005AJ....130...65C} Crowl, H.~H., Kenney, 
J.~D.~P., van Gorkom, J.~H., \& Vollmer, B.\ 2005, \aj, 130, 65 


\bibitem[Crutcher(1991)]{1991IAUS..147...61C} Crutcher, R.~M.\ 1991, 
Fragmentation of Molecular Clouds and Star Formation, 147, 61 


\bibitem[Dickey 
\& Gavazzi(1991)]{1991ApJ...373..347D} Dickey, J.~M., \& Gavazzi, G.\ 1991, \apj, 373, 347 


\bibitem[Duc 
\& Bournaud(2008)]{2008ApJ...673..787D} Duc, P.-A., \& Bournaud, F.\ 2008, \apj, 673, 787 


\bibitem[Ferland et al.(1998)]{1998PASP..110..761F} Ferland, G.~J., 
Korista, K.~T., Verner, D.~A., Ferguson, J.~W., Kingdon, J.~B., 
\& Verner, E.~M.\ 1998, \pasp, 110, 761 

\bibitem[Fujita 
\& Nagashima(1999)]{1999ApJ...516..619F} Fujita, Y., \& Nagashima, M.\ 1999, \apj, 516, 619 


\bibitem[Furlanetto et al.(2005)] Furlanetto, S.r., Schaye, J., Springel, V., \& Hernquist, L. 2005, ApJ, 622, 7

\bibitem[Gavazzi(1989)]{1989ApJ...346...59G} Gavazzi, G.\ 1989, \apj, 346, 59 

\bibitem[Gavazzi et al.(2001)]{2001ApJ...563L..23G} Gavazzi, G., Boselli, 
A., Mayer, L., Iglesias-Paramo, J., V{\'{\i}}lchez, J.~M., 
\& Carrasco, L.\ 2001, \apjl, 563, L23 

\bibitem[Gavazzi et 
al.(1995)]{1995A&A...304..325G} Gavazzi, G., Contursi, A., Carrasco, L., Boselli, A., Kennicutt, R., Scodeggio, M., \& Jaffe, W.\ 1995, \aap, 304, 325 

\bibitem[Gavazzi 
\& Jaffe(1987)]{1987A&A...186L...1G} Gavazzi, G., \& Jaffe, W.\ 1987, \aap, 186, L1 

\bibitem[Gunn 
\& Gott(1972)]{1972ApJ...176....1G} Gunn, J.~E., \& Gott, J.~R.~I.\ 1972, \apj, 176, 1 

\bibitem[Haardt 
\& Madau(2001)]{2001cghr.confE..64H} Haardt, F., \& Madau, P.\ 2001, Clusters of Galaxies and the High Redshift Universe Observed in X-rays,  

\bibitem[Hartmann(2003)]{2003ApJ...585..398H} Hartmann, L.\ 2003, \apj, 
585, 398 

\bibitem[Haynes et 
al.(1984)]{1984ARA&A..22..445H} Haynes, M.~P., Giovanelli, R., \& Chincarini, G.~L.\ 1984, \araa, 22, 445 

\bibitem[Haynes et al.(2007)]{2007ApJ...665L..19H} Haynes, M.~P., 
Giovanelli, R., \& Kent, B.~R.\ 2007, \apjl, 665, L19 

\bibitem[Hernquist(1993)]{1993ApJS...86..389H} Hernquist, L.\ 1993, \apjs, 
86, 389 

\bibitem[Irwin 
\& Sarazin(1996)]{1996ApJ...471..683I} Irwin, J.~A., \& Sarazin, C.~L.\ 1996, \apj, 471, 683 


\bibitem[Kapferer et 
al.(2009)]{2009A&A...499...87K} Kapferer, W., Sluka, C., Schindler, S., Ferrari, C., \& Ziegler, B.\ 2009, \aap, 499, 87 


\bibitem[Kenney et al.(2008)]{2008ApJ...687L..69K} Kenney, J.~D.~P., Tal, 
T., Crowl, H.~H., Feldmeier, J., \& Jacoby, G.~H.\ 2008, \apjl, 687, L69 


\bibitem[Kenney et al.(2004)]{2004AJ....127.3361K} Kenney, J.~D.~P., van 
Gorkom, J.~H., \& Vollmer, B.\ 2004, \aj, 127, 3361 


\bibitem[Kim et al.(2008)]{2008ApJ...688..931K} Kim, D.-W., Kim, E., 
Fabbiano, G., \& Trinchieri, G.\ 2008, \apj, 688, 931 


\bibitem[Koopmann et al.(2008)]{2008ApJ...682L..85K} Koopmann, R.~A., et 
al.\ 2008, \apjl, 682, L85 


\bibitem[Larson(2003)]{2003RPPh...66.1651L} Larson, R.~B.\ 2003, Reports on 
Progress in Physics, 66, 1651 

\bibitem[Larson et al.(1980)]{1980ApJ...237..692L} Larson, R.~B., Tinsley, 
B.~M., \& Caldwell, C.~N.\ 1980, \apj, 237, 692 

\bibitem[Machacek et al.(2006)]{2006ApJ...644..155M} Machacek, M., Jones, 
C., Forman, W.~R., \& Nulsen, P.\ 2006, \apj, 644, 155 

\bibitem[McKee 
\& Cowie(1977)]{1977ApJ...215..213M} McKee, C.~F., \& Cowie, L.~L.\ 1977, \apj, 215, 213 

\bibitem[Men et 
al.(2008)]{2008A&A...486..819M} Men, H., Ferri{\`e}re, K., \& Han, J.~L.\ 2008, \aap, 486, 819 


\bibitem[Miyamoto 
\& Nagai(1975)]{1975PASJ...27..533M} Miyamoto, M., \& Nagai, R.\ 1975, \pasj, 27, 533 


\bibitem[Moore et al.(1996)]{1996Natur.379..613M} Moore, B., Katz, N., 
Lake, G., Dressler, A., \& Oemler, A.\ 1996, \nat, 379, 613 

\bibitem[Moran et al.(2007)]{2007ApJ...671.1503M} 
Moran, S.~M., Ellis, R.~S., Treu, T., Smith, G.~P., Rich, R.~M.,  \& Smail, I.\ 2007,  \apj, 671, 1503 

\bibitem[Mori 
\& Burkert(2000)]{2000ApJ...538..559M} Mori, M., \& Burkert, A.\ 2000, \apj, 538, 559 

\bibitem[Norman 
\& Bryan(1999)]{1999ASSL..240...19N} Norman, M.~L., \& Bryan, G.~L.\ 1999, Numerical Astrophysics, 240, 19 

\bibitem[O'Shea et al.(2004)]{2004astro.ph..3044O} O'Shea, B.~W., Bryan, 
G., Bordner, J., Norman, M.~L., Abel, T., Harkness, R., 
\& Kritsuk, A.\ 2004, arXiv:astro-ph/0403044 

\bibitem[Oosterloo 
\& van Gorkom(2005)]{2005A&A...437L..19O} Oosterloo, T., \& van Gorkom, J.\ 2005, \aap, 437, L19 


\bibitem[Osterbrock(1989)]{1989agna.book.....O} Osterbrock, D.~E.\ 1989, 
Research supported by the University of California, John Simon Guggenheim 
Memorial Foundation, University of Minnesota, et al.~Mill Valley, CA, 
University Science Books, 1989, 422 p.,  

\bibitem[Raymond 
\& Smith(1977)]{1977ApJS...35..419R} Raymond, J.~C., \& Smith, B.~W.\ 1977, \apjs, 35, 419 

\bibitem[Roediger 
\& Br{\"u}ggen(2008)]{2008MNRAS.388..465R} Roediger, E., \& Br{\"u}ggen, M.\ 2008, \mnras, 388, 465 


\bibitem[Roediger 
\& Br{\"u}ggen(2006)]{2006MNRAS.369..567R} Roediger, E., \& Br{\"u}ggen, M.\ 2006, \mnras, 369, 567 

\bibitem[Roediger 
\& Br{\"u}ggen(2007)]{2007MNRAS.380.1399R} Roediger, E., \& Br{\"u}ggen, M.\ 2007, \mnras, 380, 1399 

\bibitem[Roediger et al.(2006)]{2006MNRAS.371..609R} Roediger, E., 
Br{\"u}ggen, M., \& Hoeft, M.\ 2006, \mnras, 371, 609 


 \bibitem[Sarazin \& White(1987)]{cooling}
Sarazin, C. \& White, 1987, Apj, 320, 32

\bibitem[Schaye(2004)]{2004ApJ...609..667S} Schaye, J.\ 2004, \apj, 609, 
667 


\bibitem[Solanes et al.(2001)]{2001ApJ...548...97S} Solanes, J.~M., 
Manrique, A., Garc{\'{\i}}a-G{\'o}mez, C., Gonz{\'a}lez-Casado, G., 
Giovanelli, R., \& Haynes, M.~P.\ 2001, \apj, 548, 97 


\bibitem[Spitzer(1978)]{1978ppim.book.....S} Spitzer, L.\ 1978, New York 
Wiley-Interscience, 1978.~333 p.,  


\bibitem[Sullivan et al.(1981)]{1981AJ.....86..919S} Sullivan, W.~T., III, 
Bates, B., Bothun, G.~D., \& Schommer, R.~A.\ 1981, \aj, 86, 919 


\bibitem[Sun et al.(2007)]{2007ApJ...671..190S} Sun, M., Donahue, M., 
\& Voit, G.~M.\ 2007, \apj, 671, 190 


\bibitem[Sun et al.(2006)]{2006ApJ...637L..81S} Sun, M., Jones, C., Forman, 
W., Nulsen, P.~E.~J., Donahue, M., \& Voit, G.~M.\ 2006, \apjl, 637, L81 


\bibitem[Sun 
\& Vikhlinin(2005)]{2005ApJ...621..718S} Sun, M., \& Vikhlinin, A.\ 2005, \apj, 621, 718 

\bibitem[Tasker 
\& Bryan(2006)]{2006ApJ...641..878T} Tasker, E.~J., \& Bryan, G.~L.\ 2006, \apj, 641, 878 


\bibitem[Trachternach et al.(2008)]{2008AJ....136.2720T} Trachternach, C., 
de Blok, W.~J.~G., Walter, F., Brinks, E., 
\& Kennicutt, R.~C.\ 2008, \aj, 136, 2720 

\bibitem[Tonnesen 
\& Bryan(2009)]{2009ApJ...694..789T} Tonnesen, S., \& Bryan, G.~L.\ 2009, \apj, 694, 789 


\bibitem[Tonnesen et al.(2007)]{2007ApJ...671.1434T} Tonnesen, S., Bryan, 
G.~L., \& van Gorkom, J.~H.\ 2007, \apj, 671, 1434 
\bibitem[Tonnesen 
\& Bryan(2008)]{2008ApJ...684L...9T} Tonnesen, S., \& Bryan, G.~L.\ 2008, \apjl, 684, L9 


\bibitem[van den Bosch et al.(2008)]{2008MNRAS.387...79V} 
van den Bosch, F.~C., Aquino, D., Yang, X., Mo, H.~J., Pasquali, A., McIntosh, D.~H., Weinmann, S.~M., 
\& Kang, X.\ 2008,  \mnras, 387, 79 

\bibitem[Vollmer et al.(2001)]{2001ApJ...561..708V} Vollmer, B., Cayatte, 
V., Balkowski, C., \& Duschl, W.~J.\ 2001, \apj, 561, 708 

\bibitem[Vollmer 
\& Huchtmeier(2007)]{2007A&A...462...93V} Vollmer, B., \& Huchtmeier, W.\ 2007, \aap, 462, 93 

\bibitem[Vollmer et 
al.(2009)]{2009A&A...496..669V} Vollmer, B., Soida, M., Chung, A., Chemin, L., Braine, J., Boselli, A., \& Beck, R.\ 2009, \aap, 496, 669 

\bibitem[Wang et al.(2004)]{2004ApJ...611..821W} Wang, Q.~D., Owen, F., 
\& Ledlow, M.\ 2004, \apj, 611, 821 

\bibitem[Warmels(1988)]{1988A&AS...72..427W} Warmels, R.~H.\ 1988, \aaps, 72, 427 


\bibitem[Yagi et al.(2007)]{2007ApJ...660.1209Y} Yagi, M., Komiyama, Y., 
Yoshida, M., Furusawa, H., Kashikawa, N., Koyama, Y., 
\& Okamura, S.\ 2007, \apj, 660, 1209 

\bibitem[Yoshida et al.(2004)]{2004AJ....127...90Y} Yoshida, M., et al.\ 
2004a, \aj, 127, 90 

\bibitem[Yoshida et al.(2004)]{2004AJ....127.3653Y} Yoshida, M., et al.\ 
2004b, \aj, 127, 3653 

\bibitem[Yoshida et al.(2008)]{2008ApJ...688..918Y} Yoshida, M., et al.\ 
2008, \apj, 688, 918 


\bibitem[Yoshida et al.(2002)]{2002ApJ...567..118Y} Yoshida, M., et al.\ 
2002, \apj, 567, 118 


\end{thebibliography}
\end{document}